\documentclass[11pt,a4paper]{article}
\pdfoutput=1

\usepackage{jheppub}

\usepackage{epsfig,multicol,bbm}
\usepackage{amsmath,amsfonts,amssymb,longtable,mathtools}
\usepackage{array}
\usepackage{graphicx}
\usepackage{subfig}
\usepackage{enumerate}

\def\beq{\begin{equation}}
\def\eeq{\end{equation}}
\def\bea{\begin{eqnarray}}
\def\eea{\end{eqnarray}}

\def\lp{\left(}
\def\rp{\right)}
\def\lb{\left[}
\def\rb{\right]}

\newcommand{\ga}{\gamma}

\newcommand{\la}{\lambda}

\newcommand{\x}{\chi}
\newcommand{\vh}{\Gamma_0}
\newcommand{\tgam}{\tilde{\gamma}}

\input epsf.sty

\preprint{NORDITA-2013-99, DCPT-13/55}

\begin{document}

\title{Dark D-brane Cosmology}

\author{Tomi Koivisto\footnote{t.s.koivisto@astro.uio.no}$^{a,b,c}$,}
\affiliation{${}^a$Institute for Theoretical Astrophysics, University of Oslo, P.O. Box 1029 Blindern, N-0315 Oslo, Norway \newline
${}^b$Nordita, KTH Royal Institute of Technology and Stockholm University, Roslagstullsbacken 23, SE-10691 Stockholm, Sweden \newline
${}^c$ Helsinki Institute of Physics, FIN-00014 University of Helsinki}
\author{Danielle Wills\footnote{d.e.wills@durham.ac.uk}$^{d}$,}
\affiliation{${}^d$Centre for Particle Theory, Department of Mathematical Sciences, Durham University, South Road, Durham, DH1 3LE, UK}
\author{Ivonne Zavala\footnote{e.i.zavala@rug.nl}$^{e}$}
\affiliation{${}^e$Centre for Theoretical Physics, University of Groningen,
Nijenborgh 4, 9747 AG Groningen, The Netherlands}

\abstract{ Disformally coupled cosmologies arise  from Dirac-Born-Infeld actions  in Type II string theories, when matter resides on a moving hidden sector D-brane. 
Since such matter interacts only very weakly with the standard model particles, this scenario can provide a natural origin for the dark sector of the universe with a clear geometrical interpretation: dark energy  is identified with the scalar field associated  to the D-brane's position  as it moves in the internal space, 
acting as quintessence, while dark matter is identified with  the matter living on the D-brane,
which can be modelled by a perfect fluid. 
The coupling functions are determined by the (warped) extra-dimensional geometry, and are thus constrained by the theory.  The resulting cosmologies are studied using both dynamical system analysis and numerics. From the dynamical system point of view, one free parameter controls the cosmological dynamics, given by the ratio of the warp factor and the potential energy scales. The disformal coupling allows for new scaling solutions that can describe accelerating cosmologies alleviating the coincidence problem of dark energy. In addition, this scenario may ameliorate the fine-tuning problem of dark energy, whose small value 
may be attained dynamically, without requiring the mass of the dark energy field to be unnaturally low. 
}

\keywords{Dark energy theory, dark matter theory, string theory and cosmology}

\maketitle

\section{Introduction}\label{introduction}

The Nobel prize-earning experimental confirmation that our universe is currently accelerating \cite{SN1,SN2} has presented a challenge for theory which remains among the most important open questions in modern cosmology. This late-time acceleration of the universe must be driven by some hitherto unidentified energy source, generally referred to as dark energy (DE). It may simply be due to a tiny cosmological constant, however more general, dynamical forms of energy are allowed by the data. Dark energy constitutes about 68\% of the overall energy content in the universe, while in the standard model of cosmology, another 27\% is in the form of cold dark matter (DM), tallying up to a total of 95\% of the energy in the universe being in the unknown dark sector. These forms of matter and energy must have little or no direct interaction with ordinary matter, as they have been observed only through their gravitational interactions.

On the other hand, fundamental theories such as string theory posit the existence of extra dimensions which can contain all kinds of matter fields, which although coexisting with our world of standard model particles in four dimensions, are spatially separated from visible matter in higher dimensions. Could it be that the mysterious dark fluids in the universe are due to the presence of another four-dimensional ``world'' which is separated from ours by additional dimensions of space?   

In addition to the dark fluids themselves, there is the question of possible interactions between these fluids. It is often simply assumed that the components of the dark sector are independent and do not interact directly, however there is no fundamental principle which forbids some form of interplay between them. Indeed, whereas new forces between DE and normal matter particles are heavily constrained by observations ({\it e.g.}~by solar system tests as well as gravitational experiments on Earth), this is not the case for DM particles. In other words, it is possible that the dark components interact with each other, while not being coupled to standard model particles. Several phenomenological interacting DE/DM models have been proposed in the literature
(see {\it e.g.}~\cite{Ukranian} for a recent review with several references), however, typically without a compelling fundamental origin for the form of the proposed couplings.

In the current work we propose a unified picture of the dark phenomena in the universe in which dark matter and dark energy are naturally interacting.
Specifically, we suggest that the cosmological dark sector, namely dark energy, dark matter and any possible dark radiation, may be naturally unified as distinct phenomena arising from the fluctuations of a single object, which we call the {\em Dark D-brane}, moving in a higher dimensional space-time  \cite{essay}.

The Dark D-brane world scenario we propose can arise from ``hidden sector branes'', which are ubiquitous in string theory D-brane constructions, which are currently moving in the six (warped) extra dimensions. Hidden sector D-branes are those branes which have no intersection with the stack of D-branes responsible for the visible sector and therefore they interact with the visible sector only gravitationally or via very massive states that are integrated out of the low energy theory. Thus the matter fields on these branes are \emph{dark} by construction. For a single D-brane, the matter fields are U(1) gauge fields which may be massive or massless. Hence they can simultaneously provide candidates for a dark matter species and a dark radiation species in the universe, where we might expect that today the dark matter species is in some form of massive decay products of these fields.
This though may not the end of the story, because if in addition, some of these hidden branes are able to move during the present epoch, other light degrees of freedom will arise which could act as dark energy and thus complete the dark spectrum in four dimensions.

In this picture, the oscillations of the open strings along the surface of a D-brane can be thought of as encoding the D-brane's \emph{intrinsic} fluctuations. These give rise to the matter fields on the brane.
Similarly, the oscillations of the open strings that are transverse to the D-brane encode the D-brane's \emph{extrinsic} fluctuations, namely the fluctuations of its position in the higher dimensions in which it is embedded. These give rise to
scalar fields which parameterise the location of the brane.
In our scenario we  associate the intrinsic fluctuations of a hidden sector brane, equivalently the open string oscillations on its surface, with dark matter or dark radiation, and its extrinsic fluctuations along a single direction in the internal six dimensional space, equivalently the open string oscillations transverse to its surface, with dark energy.
In this sense, the dark sector is unified as distinct phenomena arising from the fluctuations of the Dark D-brane in the higher dimensional spacetime today \cite{essay}.
In the current work, we consider for concreteness and to illustrate our idea, a Dark D-brane prototype in the form of a single probe D3-brane  inhabited by massive particles\footnote{For standard compactifications of Type IIB with O3/O7 orientifold planes, the U(1) gauge field on a D3-brane remains massless. However, considering the simplest Dark D-brane world scenario will be enough to illustrate our points. This can then be readily expanded upon to construct a more realistic Dark D-brane model involving higher dimensional branes and more complex D-brane configurations, where massive particles arise naturally.}.

As a geometrical framework for describing the dark sector, it is compelling that the dark fluids in this scenario turn out to be non-minimally coupled in a very particular way: the coupling, which arises due to the induced metric on the brane, is precisely a realisation of the so-called \emph{disformal transformation}  \cite{Bekenstein},
which has recently been receiving growing attention in the modified gravity literature. Indeed, it can be argued that this transformation, which takes the form
\beq\label{disformal}
\bar{g}_{\mu\nu}=C(\phi,X)g_{\mu\nu}+D(\phi,X)\partial_\mu \phi \,\partial_\nu \phi,
\eeq
is the most  general physically consistent relation between two metrics which can be given by a single scalar field $\phi$, where $X=(\partial\phi)^2$ is the kinetic term \cite{Bekenstein}. The first term in Eq.~(\ref{disformal}) is the well-known {\em conformal} transformation which characterises the Brans-Dicke class of scalar-tensor theories, for which the $f(R)$ gravity theories are a widely studied example\footnote{For the role of conformal transformations in $f(R)$ and other gravity theories, see \cite{Faraoni:1998qx,DeFelice:2010aj,Amendola:2010bk}.}. The second term is the purely \emph{disformal} contribution, which is generic in extensions of general relativity. In fact, it must appear in the Einstein frame formulation of any covariant theory involving an invariant other than $R$, or of \emph{any} more general Horndeski-type scalar-tensor theory
\footnote{See \cite{Zumalacarregui:2012us,Bettoni:2013diz,Zumalacarregui:2013pma} for the role of the disformal relation in general Horndenski classes of scalar-tensor theories.}.
Another very active area of study in which the disformal coupling makes an appearance is in the field of non-linear massive gravity theories \cite{deRham:2010kj,deRham:2010ik}.

Therefore our work makes direct contact with current theories of modified gravity and provides a fundamental origin for the disformal transformation.
Indeed, whereas in the modified gravity literature the disformal coupling is usually ``put in by hand'', we will show that starting from a consistent physical theory of quantum gravity in higher dimensions, this coupling appears naturally,
rather than  as an \emph{ad hoc} modification of Einstein's theory in four dimensions. The general relation in Eq.~(\ref{disformal}) then has a concrete interpretation as the induced metric on a probe D-brane moving in a warped higher dimensional spacetime, such that $\phi$ is the scalar  field associated to the position of the brane, and the functions $C$ and $D$ are both given in terms of  the warp factor $h$. In addition, the matter that is {\em disformally coupled} cannot be just any matter but must be the matter that is localised on the moving D-brane. In our scenario we associate $\phi$ with dark energy, and the disformally coupled matter with dark matter. Therefore, we will refer often to  it   as the Disformal Dark D-brane scenario.

Disformal couplings have a rich phenomenology and thus we can expect such matter to exhibit distinctive features. Notably, whereas the function $C$ in (\ref{disformal}) is a local scale transformation that leaves the causal structure untouched, the function $D$ affects angles and thus distorts the light cones\footnote{In relativistic MOND theories, the disformal relation is crucial in mimicking lensing by dark matter \cite{Sanders:1996wk,Bekenstein:2004ne,Skordis:2005xk,Bruneton:2006gf,Skordis:2009bf}.}. This feature has been exploited, by coupling the Standard Model electromagnetic field disformally, in varying-speed-of-light theories \cite{Magueijo:2003gj,Clayton:2001rt,Magueijo:2008sx}. Constraints on such couplings to visible sector photons have been derived from both high-precision laboratory experiments of low-energy photons \cite{Brax:2012ie} and from cosmological tests based for example on the distortion of the cosmic microwave background black-body radiation \cite{vandeBruck:2013yxa,Brax:2013nsa}. Bounds can also be placed by considering disformally coupled baryonic fluids in the radiation dominated epoch of the universe's evolution \cite{vandeBruck:2012vq}.
In the current work on the other hand, it is exclusively the dark fluids which are disformally coupled. Nevertheless, our work suggests that disformal phenomenology can in fact be viewed as a probe of extra dimensional brane movement in 
for example a string theory setting\footnote{Our proposal does not need to be restricted to string theory, but could be in principle realised in a `pure' brane world scenario \cite{Akama,Rubakov,Visser,Gibbons}.}.
To ensure that everything stays causal for all values of the field and its derivatives, constraints must usually be placed on the functional forms of $C$ and $D$  \cite{Bekenstein}. In Section \ref{GenSet2} we discuss  the  causal structure for {\em disformally interacting massive particles} (which we dub ``DIMPs'') in our Dark D-brane scenario. From the point of view of the geometrical picture, we see clearly that the causal structure in the four-dimensional disformal spacetime is in fact determined by causal dynamics in the higher dimensions, and thus we see that working within a concrete physical theory, causality arises naturally without the need to place restrictions on the functions $C$ and $D$.

In the context of string theory, Eq.~(\ref{disformal}) has  been widely exploited in cosmological applications. Indeed the so-called Dirac-Born-Infeld (DBI) inflationary scenarios \cite{Silverstein:2003hf,Alishahiha:2004eh}, for which ``slow-roll" inflation is simply the non-relativistic limit, are based on this relation, where the scalar field $\phi$ plays the role of the inflaton\footnote{For reviews on D-brane inflation see \cite{Kallosh:2007ig,Burgess:2007pz,Burgess:2011fa,Baumann:2009ni,McAllister:2007bg}.}. DBI inflation arises when the D-brane is moving in a strongly warped region, commonly referred to as a warped throat. While in the standard scenarios the D-brane moves radially in such a region, the generalisation to allow motion in all six of the compact directions in the throat has been studied, and there it was found that motion in all directions other than the radial is rapidly damped by the cosmological expansion \cite{EGTZ,Easson:2007dh,Gregory:2011cd}. Thus the system quickly converges to the single field case as in Eq.~(\ref{disformal})\footnote{Furthermore, some implications of having two different metrics have been discussed in the context of brane world scenarios in \cite{Burgess:2003tz}, where it is shown that observers living on a 3-brane may experience an induced metric which bounces without violating the weak energy condition.}.

The vast majority of D-brane inflation models deal exclusively with the scalar fields associated to the transverse degrees of freedom of the brane. Any other matter fields living on the moving brane are usually not considered: apart from the inflaton, the branes are ``empty''.
However, while such assumption can be justified during early time acceleration, 
there is in principle no reason for  it during late time cosmology. It is natural to take these matter fields into account as D-branes cannot exist independently of open strings, of which the position fields are only a subset. More importantly, these matter fields can indeed give rise to interesting cosmology, as has been shown by the relatively few studies that have considered them: In the early universe context, it has been shown that they may play the role of Wilson-line inflatons in both the warped and unwarped cases \cite{Avgoustidis:2006zp,Avgoustidis:2008zu}, or instead they may act as vector curvatons on both stationary as well as moving branes \cite{Dimopoulos:2011pe,Wills:2013yga}.

In the current work we are lead to consider DBI quintessence \cite{Martin:2008xw,Burin}  where matter on the brane is taken into account. 
While DBI-type inflation models often require fine-tuning to achieve the full 60 e-folds of accelerated expansion  which are necessary to solve the horizon and flatness problems\footnote{For example, in the simplest single field DBI models this requires an unnaturally large mass term for the inflaton \cite{Silverstein:2003hf,Alishahiha:2004eh}. For more general discussion on fine-tuning in D3-brane inflation also see \cite{Pajer:2008uy,Baumann:2007ah}.}, in the dark energy context, at most one e-fold of accelerated expansion is required, since the acceleration becomes important only at a redshift of around $z = 1$. Another attractive feature of our scenario is that since the visible sector does not feel the coupling, there is no need to consider screening mechanisms for dark energy at local scales\footnote{While for the case of the purely conformal coupling ($D=0$) various screening mechanisms have been proposed to address this problem \cite{Brax:2012gr}, it turns out that for the case of the purely disformal coupling ($C=1$, $D\neq 0$), such a mechanism is already \emph{in-built} into its very structure: The disformal coupling depends on the gradients of the scalar field, thus if the field is locally static and smooth, the coupling quite neatly disappears \cite{Noller:2012sv}. Furthermore, in non-static situations, when $D\rho \gg 1$, there is no dependence on the energy density $\rho$ in the scalar field equation, and it becomes ``unsourced'' \cite{Koivisto:2012za}. However, if the scalar couples only to hidden sector matter, then no fifth force will appear in the visible sector and so the disformal screening is redundant.}. However, for this picture to be viable, it should be able to suitably address both the coincidence problem and the fine-tuning problem of dark energy.

Now, in 
standard quintessence models, the fine-tuning problem of dark energy is translated into the unnaturally small mass required for the dynamical field driving the present acceleration. 
The coincidence problem could however be addressed by the direct coupling between the dark matter and dark energy components. In particular, in the presence of the so-called scaling solutions \cite{Copeland:1997et,Ferreira:1997hj,Amendola:1999qq}, where a fraction of the dark matter density and the dark energy density stays constant, it does not appear a surprise they are of similar magnitude today. While conformally coupled theories with a large enough coupling to address the coincidence problem typically result in unacceptably enhanced perturbation growth \cite{Koivisto:2005nr,Mota:2008nj,Bean:2008ac,Erickcek:2013oma}, the viability of much more general disformally coupled quintessence models largely remains to be explored. 
A scalar with a disformal relation has been considered in \cite{Kaloper} to drive short inflation, while in \cite{Koivisto:2008ak,Zumalacarregui:2010wj} it was considered as a quintessence field. A phenomenological model of  dark matter with a disformal coupling was studied in \cite{Koivisto:2012za,Zumalacarregui:2012us}.

We will show  that these scaling solutions do indeed arise in our Disformal Dark D-brane scenario. In addition, we will see that the scale of dark energy today depends on the current value of the brane's position field, which must go to zero as the brane approaches the tip of the throat, to which it is attracted. Thus the scale of dark energy \emph{must} be incredibly low at some epoch (without requiring its mass to be extremely low). So we can conjecture that such an epoch corresponds to the present one. Therefore, both the coincidence problem as well as the fine-tuning problem can be potentially  addressed in our scenario.

Thus our Disformal Dark D-brane world scenario,  a generalisation of coupled quintessence cosmology \cite{Ukranian,Wetterich:1994bg,Amendola:1999er,Koivisto:2005nr,Amendola:2006qi,Koivisto:2006xf,Boehmer:2008av,Li:2009sy,Honorez:2010rr,Clemson:2011an,Koivisto:2012xm,Pourtsidou:2013nha}, is  a naturally {\em unified} picture of the cosmological dark sector in which dark energy arises from the motion of a hidden sector brane in the internal (warped) space
and is \emph{disformally} coupled to the dark matter fields on its world-volume\footnote{Extra dimensional dark matter has also been proposed in the brane world context where the fluctuations of our brane give rise to ``branon'' particles \cite{Cembranos:2003mr,Cembranos:2003fu,Cembranos:2004eb}. In addition, Kaluza-Klein modes in universal extra dimensions have been widely studied as viable candidates for dark matter \cite{Servant:2002aq,Cheng:2002ej,Chialva:2012rq}, see \cite{Bergstrom:2009ib,Hooper:2007qk} for reviews.}. From a ``top-down'' point of view to coupled quintessence, the scalar and matter fields appearing in our scenario have clear geometric interpretation and their properties such as the coupling can be explicitly derived given the higher dimensional fundamental 
theory\footnote{A string inspired coupled quintessence model was presented in \cite{CPT} in terms of closed string moduli.}.

To study in detail the ensuing homogenous cosmology, we will use the method of dynamical system analysis and numerical integration. Previously these methods have been applied for DBI scalar field cosmologies in\footnote{For some other works on DBI dark energy see {\it e.g.}~\cite{Abramo:2004ji,Garousi:2004ph,Panda:2005sg,Chimento:2007es,Wei:2009xg,Chimento:2009nj,Bhadra:2012qj,Brax:2012jr,Bessada:2013mka}.} \cite{Guo:2008sz,Martin:2008xw,Burin,Ahn:2009hu,Ahn:2009xd,Copeland:2010jt}, also taking into account a phenomenological coupling\footnote{However, in that case the coupling was introduced as a phenomenological interaction term added to the conservation equations, and led to qualitatively different results from what we obtain here. It has been shown that such ad hoc couplings can even result in spurious unphysical instabilities at the perturbation level \cite{Valiviita:2008iv,Gavela:2009cy}.} \cite{Kaeonikhom:2012xr}.
We will need only two ingredients to construct realistic late time cosmologies: a matter dominated fixed point that is a saddle point, allowing for a matter dominated epoch which does not last forever, and an accelerating scaling fixed point that the universe will reach around the present epoch, allowing for acceleration to follow the matter dominated epoch in such a way that the coincidence problem of dark energy may be alleviated. Remarkably, the very simplest model we consider  contains these solutions.

The structure of the paper is as follows. In Section \ref{Sec:set-up} we discuss the string theory D-brane set-up in which the disformal coupling arises and its general implications for the physics in four-dimensions.  In Section \ref{sec:cosmology} we focus on the cosmology using dynamical systems and numerical analysis. We  conclude in Section \ref{sec:conclusions}. Some general formulas, an alternative formulation and a study of a phase space of a special case are confined to the  appendices.

\section{The general set-up}\label{Sec:set-up}

In this Section we first discuss how the disformal coupling arises from D-branes in the context of Type IIB string theory warped compactifications \cite{Giddings:2001yu,Denef:2004ze,Hebecker:2006bn}. This comprises Sec.~\ref{GenSet0}.
Then in Sec.~\ref{GenSet2} we present the set-up for DIMPs on the moving brane and discuss some general physical implications of the disformal coupling.

\subsection{Disformal coupling from moving D-branes}\label{GenSet0}

Consider a warped flux compactification of Type IIB string theory , where the higher dimensional generalisations of gauge fields, the RR-forms, $F_{n+1} = dC_n$ for $n = 0,2,4$ and their duals $n = 6,8$, as well as the NSNS-form $H_3=dB_2$ are turned on in the internal six dimensional space. These fluxes back-react on the geometry, warping it. In addition, it has been shown that they  generate a potential for most of the geometric moduli present in the compactification, which allows these moduli to be stabilised \cite{Giddings:2001yu}.

Assigning the coordinates $x^\mu$ to the noncompact dimensions, where $\mu=(0,...,3)$, and the coordinates $y^A$ to the compact dimensions, with $A=(4,...,9)$, the ten dimensional metric  takes the form
\beq \label{GMNintro}
G_{MN}dx^M dx^N = h^{-1/2}(y^A)g_{\mu\nu}dx^\mu dx^\nu + h^{1/2}(y^A)g_{AB}dy^A dy^B,
\eeq
where $g_{AB}$ is the metric of the internal six dimensional Calabi-Yau manifold, and in order to preserve Lorentz symmetry in the noncompact four dimensions, the warp factor $h$ is a function of only the internal coordinates, $h=h(y^A)$.

We now want to consider probe D$p$-branes embedded in this background. Defining the coordinates $\xi^a$ on the world-volumes of the D-branes, where $a=(0,...,p)$, we can embed them into the spacetime by the mapping $x^M(\xi^a)$. This is simply a higher dimensional generalisation of the familiar point-particle worldline in four dimensions, $x^\mu(\tau)$, where $\tau$ is usually taken to be the proper time. As spatially extended objects, D-branes will also break Lorentz symmetry, and thus should be space-filling in the noncompact dimensions. We are then free to align the four-dimensional world-volume coordinates with the four-dimensional spacetime coordinates, by choosing the static gauge  $\xi^\mu = x^\mu$. In the compact dimensions on the other hand, the D-branes will naturally tend to move about as they search for the minima of their potentials, and thus the embedding functions are kept general, $y^A(\xi^a)$.

We are interested in matter fields which are confined to the brane, as discussed in the introduction. These brane fields will naturally follow geodesics of the induced metric on the D-branes, which we denote as $\bar g_{\mu\nu}$. For a D3-brane that is moving along a single compact direction $r$ for example, this is given by
\bea\label{inducedmetric}
\bar g_{\mu\nu} = G_{MN}\partial_\mu x^M \partial_\nu x^N = h^{-1/2}(r) g_{\mu\nu}+h^{1/2}(r)\partial_\mu r\partial_\nu r\,,
\eea
where the first term arises because we are in the static gauge, and $r(x^\mu)$ is proportional to the  scalar  field associated to the brane's position  parameterising its motion in the $r$ direction. We see that the induced metric on a D-brane moving along a single direction in the compact space is precisely a realisation of the disformal relation, Eq.~(\ref{disformal}), where we can readily identify the form of the couplings in terms of the warp factor  $C(r)^{-1} = D(r) = h(r)^{1/2}$ and the scalar field with the brane's position  $\phi \propto r$.
On the other hand, the metric $g_{\mu\nu}$ describes the geometry of the bulk spacetime.

In order to see how the disformal coupling arises from the D$p$-brane action, we now look in more detail at the full action describing  its dynamics. In the Einstein frame\footnote{In D dimensions the Einstein frame and string frame are related by $G_{MN}^E = e^{-\frac{4}{D-2}\varphi}G_{MN}^s$ where $\varphi$ is the dilaton.}, the DBI action of a D$p$-brane is given by\footnote{We use the following indices for the various coordinates:
\vspace{0.2cm}\\
\begin{tabular}{l l}
\hspace{2cm}$M,N = 0,...,9$ & \hspace{1cm}for 10D coordinates \\
\hspace{2cm}$\mu,\nu = 0,...,3$ & \hspace{1cm}for 4D coordinates \\
\hspace{2cm}$A,B = 4,...,9$ & \hspace{1cm}for 6D coordinates \\
\hspace{2cm}$a,b = 0,...,p$ & \hspace{1cm}for world-volume coordinates \\
\hspace{2cm}$m,n = 4,...,p$ & \hspace{1cm}internal ($p$ - 3) world-volume coordinates \\
\hspace{2cm}$i,j = p+1,...,9$ & \hspace{1cm}internal transverse to brane  coordinates
\end{tabular}}
\beq\label{DBIaction}
S_{\rm DBI}= -\mu_p \int d^{p+1}\xi e^{\frac{(p-3)}{4}\varphi}\sqrt{-\det(\bar g_{ab}+e^{-\frac{\varphi}{2}}\mathcal{F}_{ab})}
\eeq
where
\beq
\mu_{p}=(2\pi)^{-p}(\alpha')^{-\frac{(p+1)}{2}},\hspace{1cm}T_{p}=\mu_p e^{\frac{(p-3)}{4}\varphi},
\eeq
with $T_p$ being the tension of the brane, where $\alpha'=\ell_s^2$ with $\ell_s$  the string scale and
 the vacuum expectation value of the dilaton field $\varphi$ gives the string coupling as $e^{\varphi_0}=g_s$.
The pullback of the ten dimensional metric onto the D$p$-brane world-volume takes the form (\ref{inducedmetric})
\beq\label{gamma4D}
\bar g_{\mu\nu} = G_{\mu\nu} + \frac{\partial y^i}{\partial \xi^\mu}\frac{\partial y^j}{\partial \xi^v}G_{ij}
= h^{-1/2} g_{\mu\nu} + h^{1/2}\partial_\mu y^i \partial_\nu y^j  g_{ij} ,
\eeq
for  the four dimensional components, whereas
\beq
\bar g_{mn} = \frac{\partial y^l}{\partial \xi^m}\frac{\partial y^r}{\partial \xi^n}G_{lr}.
\eeq
for the internal ones. Moreover,
$\mathcal{F}_{ab}=\mathcal{B}_{ab}+ 2\pi\alpha'F_{ab}
$ 
is the gauge invariant combination of the pullback of the NSNS 2-form  $\mathcal{B}_2$ and the field strength of the world-volume $U(1)$ gauge field.

The coupling of the brane and its world-volume fields to the bulk RR-fields is described by the Wess-Zumino (WZ) action, which is given by
\beq\label{WZ}
S_{WZ} = \mu_p \int_{\mathcal{W}_{p+1}}\sum_{n}\mathcal{C}_n \wedge e^{\mathcal{F}}
\eeq
where $\mathcal{W}_{p+1}$ is the world-volume of the brane, and $\mathcal{C}_n$ are the pullbacks of the bulk RR-$C_n$ forms to which the brane couples. In this expression, the wedge product picks out the relevant terms in the exponential.
The total action for a D$p$-brane is then given by the sum of the DBI and WZ actions, namely
\beq
S_{D_{p}}= S_{\rm DBI} + S_{\rm WZ}.
\eeq

\subsubsection{The scalar sector}
\label{sec:scalar}

The four-dimensional induced metric in Eq.~(\ref{gamma4D}), which gives the kinetic terms for the brane's position fields in the DBI action (\ref{DBIaction}), is precisely of the disformal type. Indeed, comparing with Eq.~(\ref{disformal}), the scalar field $\phi$ parameterises the variation of the position of the brane in one direction in the compact space (for example the radial direction $r^2 =\sum_i (y^i)^2$), and the functions $C$ and $D$ are given in terms of  the warp factor, which depends on the scalar field associated to the position of the brane. As we have already mentioned, while the brane can have motion in all of the transverse internal directions and the warp factor can depend on all of these directions, we are well justified to consider the single field case \cite{Easson:2007dh}.

For a D3-brane as we are interested in from now on, there are no compact coordinates, and we may define the canonically normalised position field $\phi\equiv \sqrt{T_3}r$ with corresponding warp factor $h(\phi)\equiv T_3^{-1}h(r)$, for the radial direction  $r$ in a warped throat region of the compactification.
The brane acquires a potential which is Coulomb-like in the vicinity of an anti-brane, but more generally receives a variety of contributions from ``compactification effects'' such as fluxes and other objects present  in the bulk. For the case of the D3-brane in a warped throat, these effects have been explicitly computed in \cite{Baumann:2010sx}.

Finally, the D3-brane is charged under the four-form $C_4$, which appears as the first term in (\ref{WZ}) for the case of the D3-brane. We may write this charge as $C_4 = h^{-1}\sqrt{-g}\,dx^0\wedge dx^1 \wedge dx^2 \wedge dx^3$, and thus it is given in terms of the warp factor.
Ignoring for the moment the brane gauge field, after computing the determinant in the DBI action, the scalar action for a D3-brane then takes the form
\beq\label{Sphi}
S_{\phi} = -\int d^4 x \sqrt{-g} \lb h^{-1}(\phi)\!\lp \sqrt{1+h(\phi)\partial_{\mu}\phi \partial^{\mu}\phi}-1\rp + V(\phi)\rb\,.
\eeq
This action then gives us the scalar part of the action. We now take into account the matter fields living on the brane.

\subsubsection{The matter sector}
\label{sec:matter}

Let us now focus on the kinetic terms for matter on the brane, namely the U(1) gauge field, which is encoded in the DBI action (\ref{DBIaction}) above. Matter fields that live on D-branes  naturally feel the induced metric $\bar{g}_{\mu\nu}$. Indeed, we will see in Sec.~\ref{Sec:geodesics} that their associated particles follow geodesics of $\bar{g}_{\mu\nu}$. Thus, these fields see a disformal metric. To see this concretely, we can rewrite the determinant in Eq.~(\ref{DBIaction}) as follows ($p=3$)
\beq
-\det[\bar g_{\mu\nu}+e^{-\frac{\varphi}{2}}\mathcal{F}_{\mu\nu}] = -\det[\bar g_{\mu\beta}]\det[\delta^{\beta}_\nu+e^{-\varphi/2}\bar{\mathcal{F}}^{\beta}_{\,\,\nu}],
\eeq
leading to
\beq\label{DBIgamma}
S_{\rm DBI}=
-T_3 \int d^{4}x \sqrt{-\bar g}\sqrt{\det(\delta^{\beta}_\nu+e^{-\varphi/2}\bar{\mathcal{F}}^{\beta}_{\,\,\nu})}.
\eeq
Here we have denoted $\bar{\cal F}$ to make it clear that here $\cal F$ is contracted with $\bar g_{\mu\nu}$ and not with $g_{\mu\nu}$!
On the other hand, from the point of view of $g_{\mu\nu}$, the DBI action takes the form
\beq\label{DBIgmunu}
S_{\rm DBI}= -T_3 \int d^{4}x \sqrt{-g}\,h^{-1}\sqrt{\det(\delta^{\beta}_\nu+h\,\partial^\beta y^A\partial_\nu y^B g_{AB} +e^{-\varphi/2}h^{1/2}\mathcal{F}^{\beta}_{\,\,\nu})}.
\eeq
Therefore, observers living in the background spacetime see the world-volume fields following geodesics of $g_{\mu\nu}$ but new scalar fields have appeared, namely the fields associated with the position of the brane in the compact space. In addition, the warp factor now appears in the action as the function which gives both the conformal and disformal factors, $C(\phi) \equiv (T_3 \,h(\phi))^{-1/2}$ and $D(\phi) \equiv (h(\phi)/T_3)^{1/2}$ respectively, when restricted to motion in a single direction $\phi=\sqrt{T_3 }\, r$.

Expanding the square root in the DBI action we can  rewrite (\ref{DBIgmunu}) as
\beq\label{DBIgamma2}
S_{\rm DBI}=
 -T_3 \int d^{4}x \sqrt{-\bar g} \, \left(1+\frac{e^{-\varphi/2}}{4}\bar{\mathcal{F}}^{2} + \cdots \right),
\eeq
where the first term corresponds to the kinetic term for the scalar, which  appeared in  (\ref{Sphi}) above and the dots correspond to higher order terms in $\bar{\cal F}$.

 In Type IIB string theory, vector fields can acquire masses via the familiar Higgs mechanism or via a  stringy St\"uckelberg mechanism (see appendix of \cite{Dimopoulos:2011pe} for a detailed discussion).
This stringy mechanism takes place whenever the vector field couples to a two-form field in the 4D theory. Therefore if the coupling is present, the vector will acquire a mass.
Depending upon the details of the compactification, the various two-forms which give rise to vector masses may be projected out of the spectrum: this is due to the action of objects known as orientifold planes: O-planes.
In  compactifications with O3/O7 planes, the coupling for a D3-brane vector field vanishes because the associated 2-form field is projected out of the spectrum.  This entails that  D3-brane vector fields remain massless or acquire a Higgs mass for these compactifications.
 On the other hand, vector fields on branes of lower codimension, such as wrapped D5 and D7-branes, can  acquire St\"uckelberg masses in these compactifications, because the 4D two-form to which they couple remains in the spectrum. In what follows we  consider D3-branes with pressureless, \emph{i.e.} massive particles on their world-volumes, as the simplest scenario one can build. It should be clear that our study can readily be generalised to include matter fields with pressure, or branes of lower codimension.

For a D3-brane we can then collect the vector terms into a general action of the form
\beq\label{matterS}
S_{U(1)} = - \int d^4 x \sqrt{-\bar g}\, \mathcal{L}_{U(1)}\,(\bar g_{\mu\nu}),
\eeq
where we have chosen to write the action in the disformal frame to highlight that the matter field couples to the induced metric $\bar g_{\mu\nu}$.
Above we have illustrated explicitly the case where the matter living on the brane is a vector field. However, the coupling of the induced metric will be also there for more general matter fields living on the brane. Therefore below we model a generic type of  Dark D-brane matter in terms of a coupled gas of particles, our DIMPs, which will serve to illustrate the effects of the  ``disformal" coupling.

\subsubsection{The geometry}\label{sec:geometry}

The prototype warped compactification, which is smooth all the way to the tip of the throat, is given by the compact version of  the Klebanov-Strassler geometry \cite{Klebanov:2000hb,Giddings:2001yu}. It  arises due to the presence of fluxes sourced by wrapped D3 and D5-branes, and is an exact non-singular supergravity solution. Such a geometry is rather complicated, however it features an interior region which may be approximated by the simpler adS$_5\times$ S$^5$ geometry, which corresponds to the near horizon limit of a stack of $N$ D3-branes.  This  is cut off in the infra-red which corresponds to the tip of the throat. The warp factor in this case is given by
\beq\label{adswarp}
h=\frac{\lambda_{adS}}{r^4}, \qquad \lambda_{adS} = 4\pi\alpha'^2 g_s N,
\eeq
where  $g_s N \gg1$ for the supergravity approximation to be valid, while $g_s<1$ for  string perturbation  theory approximation to hold, so that the t'Hooft coupling, $\lambda_{adS} \gg 1$. For the Klebanov-Srassler (KS) geometry, the adS$_5$ approximation breaks down near the tip of the throat. Very near the tip of the KS throat the warp factor approaches a constant value  $h\rightarrow$ const.$({\cal O}(1))$ with corrections of order $ {\cal O}(r^2)$.

In what follows we study the D-brane dynamics in the mid-throat region as well as near the tip. For the former we use the adS$_5$ approximation with the warp factor given in (\ref{adswarp}) above, and for the latter, we will simply take $h$ to be a constant. This should capture the predominant behaviour of the system in the regions of interest. Furthermore, in large-volume scenarios \cite{Balasubramanian:2005zx} the effect of the warping is washed away and thus these type of  compactifications are also explored when $h\rightarrow const.$

\subsection{Disformally Interacting Massive Particles (DIMPs)}\label{GenSet2}

To outline the essential implications of the disformal coupling for particles on a moving brane,  we now adopt a  classical point-particle description in place of the usual field theory description.
This approach can also be justified as we want to describe a fluid made of galaxies, which can be seen as point particles moving in the universe. 

We  first discuss how lengths and angles are affected by the motion of the brane; in this way we contrast the disformal coupling with the conformal coupling. We will then explain how causality is guaranteed in the spacetime that is disformally related to the spacetime on a moving brane, and give the form of the local invariant speed on the brane. Turning then to general motion on the brane, we will see that from extremising the action (\ref{DMaction}) below, particles on the brane naturally follow geodesics of the induced metric. Finally, we give the form of the energy-momentum tensor for pressureless particles on the brane and compare this to the ``bare'' energy density for standard pressureless particles. This distinction will have important consequences for cosmology, as we will later explore\footnote{In this Section we  consider expressions as they appear in the present string realisation of the disformal spacetime, for general expressions see Appendix \ref{appendixGen}.}.

Consider the effective action for massive particles evolving in a $p+1$-dimensional disformal geometry $\bar g_{\mu\nu}$.
For a D3-brane ($p=3$) as we are considering, the brane actions is entirely four-dimensional and it is  simply given by\footnote{For branes of lower codimension we  extra factors  arise from the integration over the compact directions.}
\beq\label{DMaction}
S_{\rm DDM} = -\sum^N_{i=1} \int d^4 x \ m_i \sqrt{-\bar g_{\mu\nu}\dot{x}_i^\mu \dot{x}_i^\nu} \ \delta^{(4)}(x_i(\tau)- x_i),
\eeq
where we have used that $d\tau = d^4 x\, \delta^{(4)}(x_i(\tau)- x_i)$ and the dot denotes the derivative with respect to the affine parameter $\tau$. Moreover, the disformal metric $\bar g_{\mu\nu}$ is the induced metric on the brane in Eq.~(\ref{inducedmetric}).

\subsubsection{Norms and angles}

Unlike a spacetime that is purely conformally related to the background spacetime, angles as well as lengths are distorted in the disformal spacetime. The norm of a 4-vector $a^\mu$ takes the form (where we have transformed to field variables using $T_3$)
\beq\label{norm}
\bar{g}_{\mu\nu}a^\mu a^\nu =(T_3\,h(\phi))^{-1/2} \lp a^2 + h(\phi) (\partial\phi \cdot a)^2 \rp,
\eeq
where $a^2 \equiv g_{\mu\nu}a^\mu a^\nu$.

The first term is simply a conformal transformation and the second term, which projects the four-velocity along the gradients of the scalar field in addition to giving a conformal re-scaling, is the purely disformal effect of the coupling. For a cosmological scalar field, these gradients will be time-derivatives at the level of the background.

For two 4-vectors $a^\mu$ and $b^\mu$ on a moving D-brane, the angle between their 3-vector components, becomes
\beq \label{angles}
{\cos\bar{\theta}} = \frac{a\cdot b +h(\phi)(\partial\phi \cdot a)(\partial\phi \cdot b)}{\mid a\mid\mid b\mid\sqrt{1+\frac{h(\phi)}{a^2}(\partial\phi \cdot a)^2}\sqrt{1+\frac{h(\phi)}{b^2}(\partial\phi \cdot b)^2}},
\eeq
where it is understood that the vector components are taken ({\it i.e.}~$a = a^i$, $i=1,2,3$). Therefore the angles depend on the gradients of the scalar field in an intricate way.

For the case of a slowly moving brane, the disformal lengths and angles on the world-volume approach their standard values in the background spacetime, however on a fast moving or relativistic brane, the discrepancies are enhanced, thus an observer in the background spacetime would notice lengths and angles associated with disformally coupled particles becoming more and more distorted as the brane speeds up.

\subsubsection{Causality}

Let us discuss how causality arises in this context. Firstly, motion in the higher dimensional spacetime must also obey causality, and as objects with tension or mass per unit volume, D-branes follow timelike trajectories. In particular, for a scalar field which parameterises the motion of a brane in a single compact direction, we may define a Lorentz factor
\beq\label{gam}
\gamma \equiv \frac{1}{\sqrt{1+h(\phi)g^{\mu\nu}\partial_{\mu}\phi \partial_{\nu}\phi}}
\eeq
which must always be real. In the four dimensional disformal spacetime, a necessary condition for causality is that the metric $g_{\mu\nu}$ preserves Lorentzian signature for all values of the scalar field and its derivatives; and then physical particles must follow trajectories for which $d\bar{s}^2 \leq 0$. Note that in four dimensions, there are now \emph{two} invariant speeds and indeed \emph{two} copies of the Lorentz group, one associated with the background spacetime and the other with the disformal spacetime. Writing the disformal metric as
\beq\label{disformalmet}
\bar{g}_{\mu\nu} = \frac{g_{\mu\beta}}{\sqrt{T_3 h(\phi)}} [\delta^{\beta}_{\nu}+ h(\phi)\,
  \partial^{\beta}\phi \,\partial_{\nu}\phi   ],
\eeq
we see that for a time-dependent scalar field in a cosmological background, the components are just
\beq
\bar{g}_{00} = \frac{g_{00}}{\sqrt{T_3 h(\phi)}} [1+h(\phi)\,\partial^{0}\phi \,\partial_{0}\phi]\equiv
\frac{g_{00}}{\sqrt{T_3 h(\phi)}} \, \gamma^{-2},\hspace{1cm}\bar{g}_{ij}=    \frac{g_{ij}}{\sqrt{T_3 h(\phi)}}.
\eeq
The warp factor $h>0$ always and due to causality in the higher dimensions, $\gamma^{-2}>0$ always. Therefore the signature of the disformal metric is simply given by that of the four dimensional metric $g_{\mu\nu}$, and so causality is never violated.\footnote{Note that this coincides with the standard constraints given in \cite{Bruneton:2007si} for a general disformal metric as in (\ref{disformal}), namely
\beq
C(\phi,X)>0, \hspace{1cm}C(\phi,X)+D(\phi,X)X>0,
\eeq
where $X\equiv g^{\mu\nu}\partial_{\mu}\phi \partial_{\nu}\phi $, and for our case the first condition amounts to $h>0$ and the second to $\gamma^{-2}>0$,
where $C(r)=D(r)^{-1}=h(r)^{-1/2}$. For diagonal metrics, these constraints can be deduced by writing the metric in such a form as (\ref{disformalmet}) and considering the various components. In \cite{Bruneton:2007si} it is argued that if $C$ does not depend on $X$, then the second constraint can only be met if $D$ depends on $X$. In our case, neither $C$ nor $D$ depend on $X$, and yet the second constraint is ensured \emph{dynamically} as outlined above. Even if the functional form of the coupling would allow
a sign flip of the metric, in physical set-ups that does not occur, as has been previously discussed in the contexts a disformally self-coupled field \cite{Koivisto:2008ak,Zumalacarregui:2010wj} and also disformally coupled dark matter \cite{Koivisto:2012za,Zumalacarregui:2012us}.}

To write down the local invariant speed for particles on the brane we take the special relativistic limit of Eq.~(\ref{DMaction}), using Eq.~(\ref{norm}) with $a^\mu=\dot x^\mu$, and $\tau = t$. We see that the velocities of physical particles on the brane must obey
\beq
\vec{v} \equiv \frac{d\vec{x}}{dt}\hspace{0.2cm} \leq\hspace{0.2cm} \sqrt{1 - h(\phi)\dot\phi^2} = \gamma^{-1},
\eeq
therefore if the brane is moving relativistically, causality in the disformal spacetime demands that the velocities are strongly suppressed. On the other hand, the usual causal constraint for physical particles, $\vec{v} \leq 1$, is approached in the limit that the brane is moving very slowly. Hence we see that the invariant speed for particles on the moving brane is not constant but instead becomes a dynamical quantity, determined by causal motion in the higher dimensional spacetime. Indeed, from the form of Eq.~(\ref{norm}) we see that the disformal contribution is always positive, thus a null trajectory on the brane, $\bar{a}^2 = 0$, can never correspond to a spacelike trajectory in the background spacetime, \emph{i.e.}
\beq
a^2 = -h(\phi)(\partial\phi\cdot a)^2.
\eeq
We also easily see from Eq.~(\ref{norm}) that a time-like norm on the brane can appear only more time-like to us when $h>1$. Thus null cones can only remain such or become squeezed from our point of view.

Finally, let us consider two 4-vectors $A_\mu$ and $B_\mu$ on the brane. Let us call their corresponding unit vectors $a_\mu$ and $b_\mu$, and endow the contraction by the disformal metric with a hat, so that the angle between the two vectors, as measured by an observer on the brane, is $\cos\bar{\theta}=a\,\hat{\cdot}\,b$. By inverting
the relation (\ref{angles}), we can deduce the angle between the vectors as measured by an observer in our disformally related space-time:
\beq \label{angles2}
\cos\theta = \frac{a\, \hat{\cdot}\, b - \sqrt{h(\phi)/T_3}(\partial\phi\, {\cdot}\, a)(\partial\phi\, {\cdot}\, b)}{\sqrt{1- \sqrt{h(\phi)/T_3}(\partial\phi\, {\cdot}\, a)^2}\sqrt{1- \sqrt{h(\phi)/T_3}(\partial\phi\, {\cdot}\, b)^2}}
\rightarrow \frac{a\, \hat{\cdot}\, a - \sqrt{h(\phi)/T_3}(\partial\phi\, {\cdot}\, a)^2}{1- \sqrt{h(\phi)/T_3}(\partial\phi\, {\cdot}\, a)^2}\,,
\eeq
where the second form applies if $a=b$ and again 3D components are understood.
However, the formula generalises to four-vectors as well, and then $a\, \hat{\cdot}\, a=0$ for null and $a\, \hat{\cdot}\, a=-1$ for time-like vectors. This provides another way to verify that the causality is preserved, since from the second form in (\ref{angles2}) we can immediately see that neither null nor time-like vectors in the brane space-time can appear space-like to us. Thus null cones can only remain such or become squeezed from our point of view.

Since as we have already said, we want to identify   the scalar field that parameterises the motion of the brane with dark energy, and the matter fields on the brane with dark matter,  we see that the causal behaviour of the dark matter particles is determined by the dynamics of dark energy.

\subsubsection{Geodesics}\label{Sec:geodesics}

Extremising the action (\ref{DMaction}), we see that particles on the D-brane naturally follow geodesics of the disformal metric and thus the geodesic equation becomes
\beq
\ddot{x}^\mu + \bar{\Gamma}^\mu_{\alpha\beta}\dot{x}^\alpha \dot{x}^\beta = 0,
\eeq
where the disformal Levi-Civita connection $\bar{\Gamma}^\mu_{\alpha\beta}$ is
torsion-free, and can be expressed in terms of the usual connection $\Gamma^\mu_{\alpha\beta}$ associated with $g_{\mu\nu}$ as follows:

\beq\label{connection}
\bar{\Gamma}^\mu_{\alpha\beta}=\Gamma^\mu_{\alpha\beta} - \frac{h'}{2h}\delta^\mu\!_{(\alpha} \partial_{\beta)}\phi +\frac{\gamma^2}{4} \partial^\mu\phi\left(\frac{h'}{h}g_{\alpha\beta}+4h\nabla_\alpha\nabla_\beta\phi+3h'\, \partial_\alpha \phi \partial_\beta\phi\right).
\eeq
The connection $\bar{\Gamma}^\mu_{\alpha\beta}$ is the unique connection that is metric-compatible with the induced metric $\bar{g}_{\mu\nu}$ on the moving brane.

While the extra terms in Eq.~(\ref{connection}) could in principle lead to dangerous fifth forces if visible matter follows geodesics of $\bar{g}_{\mu\nu}$, in the present work only dark matter lives on the moving brane and therefore such forces, if they arise, would not impact the visible sector directly, and are not a problem for local gravity tests.
In the above expression the conformal and disformal effects arising from (\ref{inducedmetric}) are not precisely distinguishable: the two latter terms are solely due to the disformal part, but that modifies the conformal term also by the $\gamma^2$-factor. A general expression for the connection is given in the appendix \ref{AppConnection} which allows one to unpick the various contributions.

\subsubsection{Stress energy tensor}
\label{sec:set}

Let us consider now the energy density on the brane which will be important for cosmology. The stress-energy tensor for disformally coupled matter is defined in the usual manner by
\beq
T_{\mu\nu}=-\frac{2}{\sqrt{-g}}\frac{\delta\lp- \sqrt{-\bar g}\,\mathcal{L}_{DDM}\rp}{\delta g^{\mu\nu}}.
\eeq
For the point particle action (\ref{DMaction}), the stress-energy tensor is found to be
\beq
T_{\mu\nu}=\rho\, u_\mu u_\nu\,,
\eeq
where the four velocity, normalized as $u^2=-1$, is
\beq\label{uDM}
u_\mu=\frac{\dot{x}_\mu}{\sqrt{-\dot{x}^2}}\,,
\eeq
and the energy density is given as
\beq\label{disppenergy}
\rho = \sum_i {m_i}\delta^{(4)}(x^i-x^i(\tau))\,\lp\frac{1}{T_3 \,h(\phi)}\rp^{\frac{1}{4}}\sqrt{\frac{\dot{x}^2}{g}}\lb 1- h(\phi)\lp u^\mu \partial_\mu\phi\rp^2\rb^{-\frac{1}{2}}\,.
\eeq
Comparing Eq.~(\ref{disppenergy}) with the standard expression for the energy density of pressureless matter, the ``bare" energy density,
\beq\label{bare}
\rho_b = \sum_i {m_i}\delta^{(4)}(x^i-x^i(\tau))\,\sqrt{\frac{\dot{x}^2}{g}},
\eeq
we might expect that the disformally coupled fluid behaves quite differently in cosmology to a standard pressureless fluid. This is indeed the case, as will be explored in detail in what follows. In particular, the inherent coupling of dark matter to dark energy in (\ref{disppenergy}) leads to a non-conservation of the dark matter energy density, which modifies its time evolution as the universe expands.

\section{Disformal Dark D-brane Cosmology}
\label{sec:cosmology}

In what follows, we consider the phenomenology of the disformal coupling for present day cosmological evolution. 
For the sake of clarity and in order to illustrate the effects, we focus on the simplest Dark D-brane scenario, which can be easily generalised to more complex cases, as we already mentioned.
We consider a dark sector D3-brane, which contains some type of matter that we identify with dark matter, disformally coupled to the scalar describing the radial D-brane's position today, as outlined before.
The warp factor depends only on the co-ordinate $r$, say, and we  take the prototype types of warped geometries such as adS$_5\times$S$_5$ and  a constant warped factor, mimicking the close tip region of a KS throat.
We would like to stress that while moduli fields associated to the sizes of the internal space need to be stabilised today (and preferably at early times to avoid the moduli problem \cite{dCCQR}), there is in principle no reason for  D-branes' position fields present in our universe (should we live in a stringy D-brane world scenario!) to have reached their minima today. In particular, if the associated scalar fields represent no harm in today's cosmology, there is no reason to have all D-branes sitting fixed in the internal space. We thus entertain the possibility that a moving brane today can be responsible for one or both dark sectors in the universe and study the implied phenomenology.

Firstly, in subsection \ref{sec:field} we  derive the field equations of motion for the disformally coupled components in cosmology. In subsection \ref{sec:phase} we  use  analytical methods of dynamical system analysis to study the phase space of the resulting cosmology, identifying the relevant equilibrium points of the solutions and their stability in the two directly string-motivated example geometries. The analytic considerations are in agreement with the numerical results presented in subsection \ref{sec:numerics}.
A discussion on the present value of the vacuum energy and the mass of the scalar field in our scenario is presented in  section \ref{reality}. Finally in section \ref{alternative} we explore an alternative example of  disformal DBI cosmology.

\subsection{Field equations}
\label{sec:field}

Since we are interested in cosmology, we now follow the usual effective approach and couple our probe Dark D-brane as described in section \ref{Sec:set-up} to four dimensional gravity \cite{Silverstein:2003hf}.
The total action we consider is thus
\bea\label{totalS}
S&=& \frac{1}{2\kappa^2}\int d^4 x \sqrt{-g} \,R \,-
\int d^4 x \sqrt{-g} \lb h^{-1}(\phi)\!\lp \sqrt{1+h(\phi)\partial_{\mu}\,\phi \partial^{\mu}\phi}-1\rp + V(\phi)\rb  \nonumber\\
&&\hskip7cm - \int d^4 x \sqrt{-\bar g}\, \mathcal{L}_{DDM}(\bar g_{\mu\nu})\,,
\eea
where  the first term is the ordinary four-dimensional Einstein-Hilbert action, which arises from dimensional reduction of the ten dimensional closed string sector  action, $\kappa^2 =M_P^{-2} = 8\pi G$ is the reduced Planck mass in four dimensions, which is related to the internal volume
as $M_P^2 = 2V_6^{(w)}/((2\pi)^7\alpha'^4)= M_s^2\, {\cal V}_6/((2\pi)^6\pi g_s^2)$, where
$V^{(w)}_6 = \int{d^6y\sqrt{g_6}\,h}$,  ${\cal V}_6 = V_6^{w}/\ell_s^6$ and $M_s=\ell_s^{-1}$.

This summarises the effective 4-dimensional theory derived above: the second piece which is the scalar field lagrangian, was deduced in section \ref{sec:scalar}, the disformally coupled matter lagrangian was discussed in section \ref{sec:matter}  (see Eq.~\ref{DBIgamma2}) and the geometrical sector, consisting of the usual Einstein-Hilbert term, was discussed in section \ref{sec:geometry}.

The Einstein equations derived from (\ref{totalS}) are
\beq
R_{\mu\nu}- \frac{1}{2}g_{\mu\nu} R= \kappa^2  \left( T_{\mu\nu}^{\phi} + T_{\mu\nu}^{m} \right)\,,
\eeq
where the energy momentum tensors are defined as:
\beq
T^{(\phi)}_{\mu\nu}=-\frac{2}{\sqrt{-g}}\frac{\delta\lp S_\phi \rp}{\delta g^{\mu\nu}}\,, \quad
T_{\mu\nu}=-\frac{2}{\sqrt{-g}}\frac{\delta\lp- \sqrt{-\bar g}\,\mathcal{L}_{DDM}\rp}{\delta g^{\mu\nu}}\,.
\eeq
Furthermore, the equation of motion for the scalar field becomes\footnote{Note that all equations can be easily extended to the case of a general lagrangian for the scalar field of the form $P(X,\phi)$, $X=\frac12 (\partial\phi)^2$. }:
\beq\label{phieq}
\nabla_\mu\left[\gamma \,\partial^\mu\phi \right] - V'
+\frac{\gamma}{2}\frac{h'}{h^{2}}\left(\gamma^{-1} -1\right)^2 =
-\nabla_\mu\left[h \, T^{\mu\nu}\partial_\nu \phi\right]
+ \frac{T^{\mu\nu}}{2}\left[-\frac{h'}{2 \,h}g_{\mu\nu} + \frac{h'}{2}\partial_\mu\phi \partial_\nu\phi \right] \,.
\eeq
The energy momentum tensor for pressureless matter on the brane takes the form
\beq
T_{\mu\nu}=\rho \,u_\mu u_\nu,
\eeq
where for the point particle action in (\ref{DMaction}), $u_\mu$ is given by (\ref{uDM}) and the energy density $\rho$ by (\ref{disppenergy}).
For the scalar field the energy momentum tensor turns out to be:
\bea
T_{\mu\nu}^{\phi} = P_\phi \, g_{\mu\nu} + (\rho_\phi + P_\phi) u^\phi_\mu u^\phi_\nu,\\
\eea
where
\beq
u_{\mu}^{\phi} = \frac{\partial_{\mu}\phi}{\sqrt{-\partial_\mu\phi \,
\partial^\mu\phi}}
\eeq
and we have defined
\bea\label{rhoPDBI}
 \rho_\phi = \frac{\gamma-1}{h} + V \,, \qquad
 P_{\phi} = \frac{1-\gamma^{-1}}{h} - V \,,
\eea
with $\gamma$ being the Lorentz factor for the brane's motion, given in (\ref{gam}).

Due to the non-minimal coupling, the individual conservation equations for the two energy momentum tensors are modified. The conservation equation for the full system is given in the usual fashion as $\nabla_\mu (T_\phi^{\mu\nu} + T^{\mu\nu})=0$, and we have
\beq
\nabla_\mu T_\phi^{\mu\nu}  = \left[\nabla_\mu \left(\gamma\, \partial^\mu\phi\right)
   -V'+\frac{\gamma h'}{4h^2}\left(\gamma^{-1}-1\right)^2\right]\partial^\nu\phi =
Q\,\partial^\nu\phi
\eeq
where we use (\ref{phieq}) to define
\beq
Q\equiv -\nabla_\mu\left[h \, T^{\mu\nu}\partial_\nu \phi\right]
+ \frac{h'}{4h}T^{\mu\nu}\left[-g_{\mu\nu} + h\partial_\mu\phi \partial_\nu\phi \right].
\eeq
The non-conservation coupling $Q$ is consistent with the general form for disformally coupled matter derived in \cite{Koivisto:2012za} and given in (\ref{AppQ}).

The matter sector is model in the following as a disformally coupled gas of point particles, with the stress energy tensor as given in section \ref{sec:set}.

\subsubsection{Cosmological  equations}
\label{cosmoeq}

In order to study cosmology we now  restrict to a flat  Friedmann-Lama\^itre-Robertson-Walker (FLRW) line element:
\beq
ds^2=-dt^2+a^2(t)\lp dx^2+dy^2+dz^2\rp\,.
\eeq
Since the field must be homogeneous in this background, the Lorentz $\gamma$-factor becomes
\beq\label{gammaFRW}
\gamma = \frac{1}{\sqrt{1-h\,\dot{\phi}^2}}\,.
\eeq
The Friedmann equations and the Klein-Gordon equation for the scalar field become, respectively
\bea
&& H^2 =\frac{\kappa^2}{3} \left[\rho_\phi +\rho\right]\,, \label{friedmann}\\
&& \dot H + H^2 = -\frac{\kappa^2}{6}\left[ \rho_\phi+ 3P_\phi +\rho \right]\,,\\
&& \ddot \phi + \frac{h'}{2h^2}(1-3\gamma^{-2}+2\gamma^{-3}) + \gamma^{-3}(V'+Q_0)+3H\gamma^{-2}\dot\phi =0 \label{kg}\,.
\eea
We further have the continuity equation for the scalar field and matter
\bea \label{cont}
\dot\rho_\phi + 3H(\rho_{\phi}+P_{\phi}) = -Q_0\dot\phi\,, \qquad \dot\rho + 3H\rho = Q_0\dot\phi\,.
\eea
Finally, the non-conservation coupling for the background, $Q_0$, is given by
\beq \label{qq1}
Q_0 = h\rho\left[\frac{3h'}{4h}\dot\phi^2 - \frac{h'}{4h^2}+\dot\phi\left(3H + \frac{\dot\rho}{\rho}\right)+\ddot\phi\right].
\eeq
Solving away the leading derivative terms for $\phi$ and $\rho$ using Eqs.~(\ref{kg}), (\ref{cont}), this becomes
\beq
Q_0 = -\left[ \frac{h\lp V'+3\ga H\dot{\phi}\rp +\frac{h'}{h}\lp 1-\frac{3}{4}\ga\rp}{\ga+h\rho}\right]\, \rho\,.
\eeq
Let us now consider some implications of this  coupling.

\subsubsection{The effects of the coupling} \label{crosscheck}

In order to gain some understanding of the interaction between dark matter and dark energy, we can compute $Q_0$ in an alternative way.  In an FRW background, the energy density for pressureless particles on the brane given in (\ref{disppenergy}) is:

\beq\label{DMrho2}
\rho = (T_3 h)^{-1/4}\,\rho_b\,\gamma \,,
\eeq
where the bare energy density $\rho_b$ solves the standard continuity equation for uncoupled matter yielding
\beq\label{rhob2}
\rho_b = \rho_{0}\,a^{-3}.
\eeq
Taking the first derivative of (\ref{DMrho2}) using (\ref{rhob2}), we obtain the equation
\beq\label{derivrho}
\frac{\dot\rho}{\rho} +3H = \frac{\dot\gamma}{\gamma}-\frac{h'}{4h}\dot\phi\,,
\eeq
which exactly matches the second of Eqs.~(\ref{cont}) with $Q_0$ defined by (\ref{qq1}). This also allows us to write $Q_0$ in a particularly compact form
\beq \label{compact}
\frac{Q_0}{\rho}\,\dot\phi= \frac{d}{dt}\log{\lp\frac{\gamma}{h^{1/4}}\rp}.
\eeq
In conformally coupled theories, the bare energy density is modified by a field dependent conformal factor \cite{Amendola:1999er}: we see quite neatly here that new disformal effect is simply to modulate the bare energy density by an additional factor $\gamma$ that involves the kinetic term of field as well.

We can gain some additional useful insight into the dynamics of the system by rewriting (\ref{derivrho}) in terms of an effective equation of state  for the {\em disformal dark matter} (DDM),
\beq\label{weff}
\frac{\dot\rho}{\rho} + 3H(1 + w^{eff}_{DDM})=0,\qquad w^{eff}_{DDM}\equiv -\frac{1}{3H}\left(\frac{\dot\gamma}{\gamma}-\frac{h'}{4h}\dot\phi\right)= -\frac{1}{3H}\frac{Q_0\dot\phi}{\rho}\,.
\eeq
The effective equation of state simply quantifies how the dark matter dilutes with the expansion. 
In particular we see from here clearly that if $w^{eff}_{DDM}<0$ the dark matter will redshift slower than $a^{-3}$ and faster in the opposite case, $w^{eff}_{DDM}>0$. 

In a completely analogous way, we can consider an effective equation of state for the DBI field,
\beq\label{weffsc}
\frac{\dot\rho_\phi}{\rho_\phi} + 3H(1 + w_\phi^{eff})=0,\qquad w^{eff}_\phi \equiv w_\phi -\frac{\rho}{\rho_\phi}w^{eff}_{DDM}, \qquad w_{\phi}\equiv \frac{P_\phi}{\rho_\phi}\,.
\eeq
If the energy in dark matter is boosted such that $w^{eff}_{DDM}<0$, $w_\phi^{eff}$ will correspondingly receive a positive contribution from the coupling term having less accelerating power. On the other hand, if $w^{eff}_{DDM}>0$ then dark energy is draining energy from dark matter and thus $w_\phi^{eff}$ receives a negative contribution from the coupling term having more accelerating power.

As can be seen from (\ref{weff}), the sign of the effective equations of state depends on the behaviour of $\dot\gamma$, $h'$ and $\dot \phi$. Note first that $\dot\gamma$ will always start being positive as the brane starts moving down the throat. In the case of a smooth throat such as the KS one, the brane will eventually start slowing down till $\gamma\to 0$.
Now, the warp factor is always positive and it grows as we reach the tip of the throat at $\phi=0$, therefore the contribution from the warp factor is always negative. On the other hand, the sign of $\dot\phi$ depends on whether the brane is moving down or up the throat.

Let us consider more explicitly the case we will be mostly interested in, an adS-like throat, such that $h\propto \phi^{-4}$. In this case the time-dependent combination $h^{-1/4}\gamma$ which appears in 
(\ref{compact}) is simply $\phi\gamma$. The general solution to the continuity equation in Eq.~(\ref{weff}) is then $\rho \propto a^{-3(1+w_{DDM}^{eff})}$. Due to the fact that $\rho_b \propto a^{-3}$, we see that $\phi\gamma \propto a^{-3w_{DDM}^{eff}}$ in Eq.~(\ref{DMrho2}). Thus the disformal coupling may either quicken or slow the dilution of dark matter, depending on whether $w_{DDM}^{eff}$ is positive or negative, as explained above. Now, despite the fact that $\phi \rightarrow 0$ as the brane moves towards the tip of the throat, the Lorentz factor may in fact grow rapidly enough such that the overall effect is that $\phi\gamma$ is growing with time. This would imply that $w_{DDM}^{eff}$ in Eq.~(\ref{weff}) becomes \emph{negative} due to the presence of the coupling term,
so that the dark matter energy density dilutes slower with the expansion due to the energy interchange with dark energy. So, interestingly, while the the conformal contribution $\sim d\log{\phi}/dt$ in Eq.~(\ref{weff}) tends to quicken the dilution of dark matter particles, the disformal effect $\sim d\log{\gamma}/dt$ acts against the dilution and could, if it dominates, serve to boost the energy density residing in dark matter.

In the following it will be useful to define the total equation of state $w$ which characterises the expansion rate as
\beq \label{t_eos}
w_T \equiv -\frac{2\dot{H}}{3H^2}-1 = \frac{P_\phi}{\rho_\phi + \rho}.
\eeq
This is the quantity that is relevant for observations, and is the ratio of the total pressure content of the universe and its total energy density.

In a so-called \emph{scaling solution}, where $w_{\phi}^{eff}=w_{DDM}^{eff}$,  the scaling components  dilute at the same rate and their fractional energy densities maintain a constant ratio. In this case, $w_{\phi/DDM}^{eff}=P_\phi/(\rho_\phi + \rho) = w_T$ as we see from the second equation in (\ref{weffsc}). \emph{Accelerating} scaling solutions then occur when $w_{\phi}^{eff}=w_{DDM}^{eff} < -1/3$. In the following we will show, using the method of dynamical system analysis, that such solutions arise for the model at hand due to the presence of the disformal coupling. Obviously, in the absence of the coupling, there can be no accelerating scaling solutions, since then $w_{DDM}^{eff}=w_{CDM}=0$ as seen in Eq. (\ref{weff}).

In a DBI scenario where matter on the brane is not taken into account, $w_T\equiv w_\phi$ becomes negative when the brane is relativistic and warping is strong, because in that case the pressure $p_\phi \rightarrow -V$ in (\ref{rhoPDBI}), and acceleration is attained when $w_\phi < -1/3$. 
Expansion is of the power law type, while quasi de Sitter expansion, for which $w\sim-1$, can usually only arise in the slow-roll limit of DBI, for then $\gamma\sim1$ and so $\rho_\phi\sim V \sim -P_\phi$ in Eqs.~(\ref{rhoPDBI}). In this case the DBI dark energy field is almost constant as the universe expands.

On the other hand, in the present case where matter on the brane is taken into account, $w_T$ is given by Eq.~(\ref{t_eos}), which reduces to $w_\phi$ only when matter is diluted away and thus $\rho \rightarrow 0$. One consequence of this is that if $w_{DDM}^{eff}$ is positive, $w_\phi^{eff}$ can be pushed to $-1$ even when
 $w_\phi>-1$, \emph{i.e.}, dark energy can be constant as the universe expands. This occurs if the purely conformal contribution $\sim d\log{\phi}/dt$ in Eq.~(\ref{weff}) is dominating over the disformal contribution $\sim d\log{\gamma}/dt$. As mentioned above, the other possibility is if the growth of the Lorentz factor is rapid enough, the disformal effect can dominate and thus the energy density in matter can be boosted, resulting in $w_{DDM}^{eff}$ becoming negative. For a brane moving towards the tip of a warped throat, we expect that the Lorentz factor will grow very rapidly at first. This could result in $w_{DDM}^{eff}<-1/3$, \emph{i.e.} the disformally coupled matter could contribute to driving the expansion of the universe: this is the emergence of a scaling solution. Then, once the strong warping forces the growth of $\gamma$ to become less rapid, the disformal dark matter will be less boosted and might be diluted away, eventually giving rise to a standard DBI epoch.

So here we see the emergence of a new aspect to the usual DBI scenarios: the Lorentz factor acts on $\rho_b$ to slow down its usual dilution by $a^{-3}$, allowing for the possibility of a new epoch of accelerated expansion that is driven in part by the disformally coupled dark matter on the brane. This could eventually evolve into the standard scenario in which matter does not contribute to the expansion. In this way, the accelerated expansion of the current universe can begin in a matter dominated era, during which the disformally coupled dark matter fluid is active in initiating the acceleration of the expansion, due to its non-minimal coupling to dark energy. This early accelerating era featuring an interplay between dark matter and dark energy eventually gives way to a fully dark energy dominated era. In what follows, these various regimes will be explored using both dynamical systems analysis as well as numerical examples.

\subsection{Phase space analysis}
\label{sec:phase}

In this section we  make use of a dynamical system approach to solve the equations of motion. In this approach
one considers a system of coupled differential equations put into to first order form:
\beq
\dot{\bf x} = {\bf f}({\bf x})\,.
\eeq
The vector ${\bf x}$ may have any integer dimension and its components  span the phase space of the same dimension. The evolution of the system is described by trajectories in this phase space. The fixed points (or equilibrium points or critical points as they are sometimes called), are those points in the phase space where the trajectories may stay constant. At a fixed point ${\bf x} = {\bf x}_c$ then,
\beq
\dot{\bf x}_c = {\bf f}({\bf x}_c) = 0\,.
\eeq
For example, if one of the components of ${\bf x}$ was $H$, all the fixed points would correspond to de Sitter solutions with different constants $H=H_c$. This illustrates two basic points: the fixed points do not need to describe static situations (in the de Sitter example the scale factor is evolving with time), and secondly how one sets up the phase space, {\it i.e.}~chooses the variable combinations that define ${\bf x}$, determines very crucially whether the fixed points correspond to interesting situations of the system at hand or not (if one would have chosen $aH$ as the variable instead of $H$, the fixed points would correspond to turnarounds or trivial solutions instead of de Sitter ones). The linear stability of each fixed point 
with respect to small perturbations defined by ${\bf x} = {\bf x}_c + {\bf \delta x}$, can be studied from the first order perturbed equations:
\beq
 \dot{\bf x} =  {\bf f}({\bf x}_c + {\bf \delta x}) =  {\bf f}({\bf x}_c) + {\bf F}\cdot {\bf \delta x} + \cdots \Rightarrow \dot{\bf \delta x} =  {\bf F}\cdot {\bf \delta x} \,,
 \eeq
where ${\bf F}$ is a matrix with the components $F_{MN}=\partial f_M/\partial x_N$ and the equality holds up to the linear order. Now, in an orthogonal basis, ${\bf F}^{(o)}$ is just a diagonal matrix consisting of the eigenvalues and the above equation has the  solution  $\delta x_N^{(o)} \sim \exp{(\lambda_N)}$, where $\lambda_N$ is the eigenvalue corresponding to the orthogonal basis vector $\delta x_N^{(o)}$.
Since the eigenvalues are independent of the basis, we can compute them directly from ${\bf F}$.
Stability is then determined from the eigenvalues as follows: a) if all the $\lambda_N<0$ are negative, all the perturbations decay, that is the fixed point is stable  and we call it an {\it attractor}. b) If all $\lambda_N>0$ are positive, any fluctuation away from the fixed point will grow and take the system away from the solution ${\bf x} = {\bf x}_c$. This point is thus unstable and can  be called a {\it repellor}. c) Finally, if some of the eigenvalues are positive and at least one negative, the system is unstable when disturbed in some direction in the phase space, while stable to disturbances in some other directions. Such a solution is called a {\it saddle point}.

Let us then consider our specific system of cosmological equations in section \ref{cosmoeq}. The phase space is there two-dimensional: we can formulate the equations as three coupled first order differential equations (for the dark energy scalar field, its derivative, and the matter density), subject to one constraint (given by the Friedmann equation). A convenient choice of variables turns out to be 
\beq
x \equiv \frac{\kappa \gamma}{\sqrt{3\lp \gamma +1\rp}}\frac{\dot{\phi}}{H}\,, \quad  z \equiv \frac{\kappa\sqrt{V}}{\sqrt{3}H}\,, \quad \Omega \equiv \frac{\kappa^2\rho}{3H^2}\,.
\eeq
The Friedmann constraint (\ref{friedmann}) allows us to then to eliminate $\Omega$ as
\beq \label{friedmann2}
\Omega = 1- x^2 - z^2\,,
\eeq
leaving us with physical space spanned by\footnote{We assume positive potential energies for the field in the following. However the formulas would apply also for negative potentials when extended to imaginary $z$.} $-1\le x \le 1$, $0 \le z \le 1$. Furthermore, it is convenient to use, instead of $\gamma$, the variable
\beq
\tilde{\gamma} \equiv \frac{1}{\gamma}\,, \quad 0 \leq \tilde{\gamma} \leq 1\,.
 \eeq
The expansion rate corresponding to each parameter value is described then by the total equation of state defined in (\ref{t_eos}),
\beq \label{t_eos2}
w_T=\tilde{\ga}x^2 -z^2\,.
\eeq
In fact, to close the system of equations, we will need also $\tgam$, and in this sense we have a three-dimensional phase space. However, as seen from (\ref{friedmann2}), it is only $x$ (roughly speaking, the kinetic energy contribution) and  $z$ (the potential energy contribution), that determine the expansion rate. For this reason, it is useful to view the phase space in terms of these variables considering $\tgam$ as a parameter. This kind
of approach was used also in Refs. \cite{Ahn:2009hu,Ahn:2009xd}.

After some algebra, using the equations of the previous section together with the definitions above, the evolution equations for the three dimensionless variables, in terms of the e-folding time $N=\log{a}$, can be brought into the following form:
\bea
&&\hskip-0.9cm\frac{dx }{dN} =  \frac{3x}{2}  \left(\frac{(\tgam+1) (2 \tgam-1) x^2}{\tgam \left(x^2+z^2-1\right)-z^2+1}+\tgam x^2-z^2+1\right) \nonumber \\
&& + \frac{\sqrt{3\tgam(\tgam+1)} x^2 \left[\mu  \left((10 \tgam-3) x^2-2 \tgam+3\right)+z^2 ((4-8 \tgam) \lambda +(2 \tgam-3) \mu )\right]}{8
   \left(\tgam \left(x^2+z^2-1\right)-z^2+1\right)}\,, \label{dxdn}
\\
&&\hskip-0.9cm\frac{dz }{dN} =  \frac{z}{2}  \left(3 + 3 \tgam x^2-3 z^2 - \sqrt{3\tgam(\tgam+1)} \lambda \, x\right)\,, \label{dzdn} \\
&&\hskip-0.9cm\frac{d\tgam }{dN}  =  \frac{3 \tgam \left(1-\tgam^2\right) x^2}{\tgam \left(x^2+z^2-1\right)-z^2+1} + \frac{\sqrt{3\tgam(\tgam+1)} (1-\tgam) \tgam x \left(\mu +3 \mu  x^2-z^2 (4 \lambda +\mu )\right)}{4 \left(\tgam
   \left(x^2+z^2-1\right)-z^2+1\right)} \label{dgdn} \,.
\eea
To close the system, we need to to express the derivatives the warp factor and the potential appearing in the field equations. For this purpose we have define the following quantities here:
\beq
\mu \equiv -\frac{h'}{\kappa h}\,, \quad
\la \equiv   -\frac{V'}{\kappa V}\,.
\eeq
In general, their evolution equations are
\bea
\frac{d\mu}{dN}   & = & \Gamma_\mu \sqrt{3\tgam(1+\tgam)} \,, \quad \Gamma_\mu = \frac{{h'}^2-h''h}{\kappa^2 h^2}\,, \\
\frac{d\la}{dN} & = & \Gamma_\lambda \sqrt{3\tgam(1+\tgam)} \,, \quad \Gamma_\la = \frac{{V'}^2-V''V}{\kappa^2 V^2}\,.
\eea
Our system can be closed, {\it i.e.}~put into an autonomous form when the two $\Gamma$-factors can be expressed in terms of the other quantities. If the functions were exponential, the $\Gamma$-factors would vanish and thus both $\mu$ and $\la$ would be constants. However, this simplification is not motivated by the geometries discussed above in Section \ref{sec:geometry}, though it might arise in some suitable brane-world scenario. We briefly check the contents of the phase space with constant $\lambda$ and $\mu$ in  appendix \ref{expos}. In the following we study the case in which the warp factor and the potential both have a power-law form. Then we will scrutinise the cases in which the powers are those corresponding to an adS$_5$ and  constant-warped geometries.

\subsubsection{Power-law evolution}

In this section, we  take  power law forms for the warp factor and the potential, which include adS$_5$ and a mass term potential:
\beq \label{forms}
h(\phi)=h_0\frac{\kappa^{4-m}}{\phi^m}\,, \quad V(\phi) = V_0 \kappa^{n-4}\phi^n\,.
\eeq
The parameters $V_0$ and $h_0$ are dimensionless numbers and $n$ and $m$ are constants. A restriction we need to impose is that $n\neq m$; then we can solve $\mu$ and $\lambda$ in terms of the other variables as
\beq
\mu = m\lp \frac{(1-\tgam) z^2}{\tgam\vh x^2}\rp^\frac{1}{m-n}\,, \quad
\lambda = -n\lp \frac{(1-\tgam) z^2}{\tgam\vh x^2}\rp^\frac{1}{m-n}\,,
\eeq
where we have defined
\beq \label{vh}
\vh \equiv h_0 V_0\,,
\eeq
which turns out to be a very useful parameter. Here and in the following we take $\phi\ge0$, 
since the field corresponds to the brane's position in the internal space.

Using the equations (\ref{dxdn}, \ref{dzdn}, \ref{dgdn}) above, together with the definitions (\ref{forms}, \ref{vh}), the evolution equations for the three dimensionless phase space variables, in terms of the e-folding time $N=\log{a}$, turn out as
\bea
&&\hskip-0.4cm
\frac{dx }{dN}   \label{sys1} =  \frac{x}{8 \left(\tgam \left(x^2+z^2-1\right)-z^2+1\right)}
\Bigg[\!
12 \!\left(\!\tgam^2 x^4 \!+\!\tgam x^2 \!\left((\tgam-2) z^2\!+\!\tgam\!+\!3\right)\!-\!\lp
\tgam-\!1\rp \!\!\left(z^2-1\right)^2
\!-\!x^2\!\right) \nonumber \\
&& +
\sqrt{3\tgam (\tgam+1)}  \,x \!\left(z^2
   (2 \tgam m+8 \tgam n-3 m-4 n)\!+m \!\left((10 \tgam-3) x^2-2 \tgam \!+\!3\right)\!\right)\!\! \left(\! \frac{\vh \tgam  x^2}{(1-\tgam)
   z^2}\!\right)^{\frac{1}{n-m}}\!\!\Bigg]\,,  \nonumber \\ \\
&&\hskip-0.4cm\label{sys2}
 \frac{dz }{dN}   =  \frac{z}{2}  \left[\sqrt{3\tgam\lp \tgam+1\rp} \,n \, x \left(\frac{\tgam \vh x^2}{(1-\tgam)
   z^2}\right)^{\frac{1}{n-m}}+3 \,\tgam \,x^2-3 z^2+3\right]\,, \\
&&\hskip-0.4cm 
\frac{d\tgam}{dN} = \! \frac{\tgam(1-\tgam) \,x \!\left[\! \sqrt{3\tgam\lp \tgam+1\rp} \left(3 \,m \,x^2-m \,z^2+m+4 n z^2\right) \left(\frac{\tgam \vh x^2}{(1-\tgam)
   z^2}\right)^{\frac{1}{n-m}}+12  (\tgam+1)  x  \right]}{4 \left(\tgam
   \left(x^2+z^2-1\right)-z^2+1\right)}\label{dtildegam}. \nonumber \\
\eea
Interestingly, the structure of the phase space depends solely upon the product of the parameters defined in ({\ref{forms}) which quantifies the energy scales of the potential and the warp factor, $\vh$ defined in equation (\ref{vh}).

\smallskip

\subsubsection*{General behaviour}

Before looking into particular solutions for the powers in the warp factor  and potential, we can make some general statements about the possible fixed points of (\ref{sys1})-(\ref{dtildegam}).
Indeed, the equations (\ref{sys1})=(\ref{sys2})=(\ref{dtildegam})=0, can be solved in various ways:

\begin{enumerate}[a)]

\item {\it Standard matter dominated solution}: in this case $x_{MD}=z_{MD}=0$ $\Rightarrow \omega_{MD} =0$ and $\Omega_{MD} =1$. This is valid for all values of $n, m$ and $\gamma$.
The 
 eigenvalues corresponding to small perturbations around this solution are $(\frac{3}{2},\frac{3}{2})$. Therefore this solution is always unstable.

\item {\it Potential dominated de Sitter solution}: in this case $x_{dS}=\Omega_{dS}=0$ so that $z_{dS}=1$ and thus  $\omega_{dS}=-1$. From (\ref{sys1}) and (\ref{sys2}) one can check that in order for it to be a solution, $n$ and $m$ must satisfy:
\beq
{n-m} \leq -2\,,
\eeq
and  the  solutions have $\tilde\gamma_{dS} =1$. The eigenvalues corresponding to this solution are $(-3,3)$,
and therefore the solution  is always a saddle point.

\item {\it Kinetic dominated solution}: in this case $x_{kin}=\pm 1$, $z_{kin}= \Omega_{kin} =0$. From  (\ref{sys1}) and (\ref{sys2}) we see that in order to be a solution to these equations we need
\beq
n-m<0
\eeq
and therefore the only solutions have $\tilde\gamma =\omega_{kin}=0$.
The eigenvalues corresponding to this solution are $3(\frac{1}{2},-1)$, thus this solution is a saddle point.

\item For more general solutions with $0<z<1$, which will be the most interesting ones, we can make some general statements and will look into two concrete examples below.
Solving (\ref{sys2}) =0 gives rise to the following equation:
\beq\label{zgral}
3(1+\tilde\gamma \,x^2)\,z^{\frac{2}{n-m}} -3\,z^{2\lp1+\frac{1}{n-m}\rp} + n\,\Gamma_0^{\frac{1}{n-m}} \sqrt{3(\tilde\gamma+1)\,\tilde\gamma}\,\,
\frac{x\lp\tilde\gamma\, x^2\rp^{\frac{1}{n-m}}  }{(1-\tilde\gamma)^{\frac{1}{n-m}}}=0 \,.
\eeq
One can check that if the last term in this equation vanishes, then we are back at one of the previous solutions. Therefore, non-trivial solutions arise when the last term does not vanish. We can then have the following situations
\begin{enumerate}[i)]
\item $x=0$ or $\tilde\gamma=0$. This case requires that
\beq
n-m =-2
\eeq
and the solution to (\ref{zgral}) can be easily found (see below).

\item For general values of $x$ and $\tilde\gamma$, the solution to (\ref{zgral}) is more complicated  depending  on the precise values of $n$ and $m$.

\end{enumerate}

\end{enumerate}

In what follows we  consider two explicit examples of the classes of solutions above, corresponding to a brane moving down an adS$_5$ throat  in a mass term potential and a constant warp factor, with an inverse law potential, where the brane moves towards the bulk geometry. This latter case can be seen as an example of a moving brane in a  large volume scenario.

\subsubsection{The adS$_5$ warp factor}
\label{ads}

Let us consider first the adS case where the warp factor goes like  $h \sim \phi^{-4}$. For the potential we consider a  mass term, that is we set  $m=4$, $n=2$, so $n-m=-2$. 
From the general discussion above we see that the system contains  classes (a), (b), (c) of fixed points. Furthermore, within class (d) we have the following fixed points:

\paragraph{Class (d):  $0<z<1$.}
In this example, the condition $dz/dN=0$  from Eq.~(\ref{zgral}) reduces to
\beq
  3\lp 1 + \tgam x^2 - z^2 \rp + 2 \, z \,\sqrt{\frac{3(1-\tgam^2)}{\vh}} S(x) = 0\,,
\eeq
where $S(x)=$ sign$(x)$. The solutions to this equation are
\bea \label{zatt}
z_\pm & = &  \frac{1}{\sqrt{3\vh}}\lb S(x)  \sqrt{1-\tgam^2} \pm \sqrt{1-\tgam^2+3\vh\lp1+ \tgam x^2\rp}\rb\,.
\eea
Thus we see that physical solutions exist only when the field is rolling down the throat, $S(x)=-1$, for the positive branch since otherwise either $z<0$ or the matter energy density is negative since $z>1$.
It is difficult to find the most general solution for $x$. However, we can focus on the  special case  $\tilde{\gamma}=0$, corresponding to an ultra relativistic regime, (d)(i) above.
In this case we obtain the following  fixed points:

\begin{itemize}
\item Matter scaling solution with $x_{DDM}=0$. For this solution we have
\beq \label{uv1}
x_{DDM}=0\,, \quad z_{DDM} = \frac{-1+\sqrt{1+3\vh}}{\sqrt{3\vh}}\,, \quad \Omega_{DDM} = \frac{2}{1+\sqrt{1+3\vh}}\,.
\eeq
The total equation of state parameter approaches minus unity as one increases $\vh$,
\beq  \label{wuv1}
w_{DDM} = -\frac{\lp 1-\sqrt{1+3\vh}\rp^2}{3\vh}\,.
\eeq
The eigenvalues for this fixed points are
\beq
\lp -\frac{1+3\vh-\sqrt{1+3\vh}}{\vh}\,, \frac{3\lp 1+\vh - \sqrt{1+3\vh}\rp}{2\vh}\rp\,.
\eeq
Requiring these to be negative, we find that this solution is stable for $0<\vh<1$. Otherwise it is a saddle point.
Moreover, we are interested in accelerating solutions, which means that the total effective equation of state (\ref{wuv1}) for this solution should satisfy $w_{DDM} < -1/3$. This  requires $\vh>1$. Therefore we see that the solution is not an accelerating attractor for  $0<\vh<1$,
however it  could be a viable matter scaling attractor when $\vh$ is small enough, such that $w \sim 0$. The reason this needs to be small is that large-scale structure would be too different from the $\Lambda$CDM case if dark matter was not effectively nearly pressureless during the structure formation era.
As we will see in the numerical study below, this fixed point is typically reached as an intermediate stage in a
 cosmic evolution which is close to $\Lambda$CDM cosmology.

\item Kinetic solution with $\Omega_{DBI}=0$.
For this ultra-relativistic solution the matter contribution vanishes and
\beq \label{uv2}
x_{DBI}= -\sqrt{\frac{2}{1+\sqrt{1+3\vh}}} \,, \quad z_{DBI} = \frac{-1+\sqrt{1+3\vh}}{\sqrt{3\vh}}\,, \quad \Omega_{DBI} = 0 \,.
\eeq
The total equation of state is the same as for the matter-scaling solution above,
\beq \label{wuv2}
w_{DBI} = -\frac{\lp 1-\sqrt{1+3\vh}\rp^2}{3\vh}\,.
\eeq
Now we obtain for the eigenvalues of this fixed points: 
\beq
\lp -\frac{1+3\vh-\sqrt{1+3\vh}}{\vh}\,, \frac{3\lp 2-\sqrt{1+3\vh}\rp }{1 + \sqrt{1+3\vh}}\rp\,.
\eeq
From this we see that this solution is a saddle point when the previous one is an attractor, that is when $0<\vh<1$.
Moreover, when $\vh>1$, the solution is an accelerating attractor with $w_{DBI} < -1/3$, while the previous one is a saddle point for these values of $\vh$.
\end{itemize}

In summary, for the adS case with a mass term we have found two accelerating attractors for the system: a nonrelativistic potential-dominated de Sitter solution (class (c)), and an ultra-relativistic DBI solution (class (d)-(i)).

\subsubsection{Constant warp factor }

In regions both asymptotically far in the bulk and very near to the tip of a Klebanov-Strassler throat, the warp factor can be approximated by a constant. This provides the simplest example of a nontrivial disformal relation, where both $C$ and $D$ are constants.
In this case   $m=0$. Following the general discussion above, we know that this case has class (a) of fixed points. Furthermore, it posses an accelerated saddle point, class (b) of solutions only for $n=-2$, that is, an inverse power law potential.  Moreover, for  all $n<0$ it posses class (c) of fixed points as well.
Regarding class (d), we have the following fixed points:

\paragraph{Class (d):  $0<z<1$.}  Focusing again in class (d)-(i), we need $n=-2$. Then Eq. (\ref{zgral}) yields for $z$
\bea \label{zatt2}
z_\pm & = &  \frac{1}{\sqrt{3\vh}}\lb -S(x)  \sqrt{1-\tgam^2} \pm \sqrt{1-\tgam^2+3\vh\lp1+ \tgam x^2\rp}\rb\,.
\eea
From here we can see that now the physical solutions correspond to a brane moving towards the bulk geometry, that is $S(x) = +1$ and furthermore we should pick the $+$-branch of the solution such that $z>0$. Focusing again in the ultra-relativistic limit $\tilde{\gamma}=0$ we  consider the cases when either the matter contribution or the kinetic contribution to the expansion are negligible.
\begin{itemize}
\item Matter scaling solution $x_{DDM}=0$. This fixed point and its total equation of state are given by the expressions (\ref{uv1}) and (\ref{wuv1}).
However, now the eigenvalues of the perturbation matrix turn out to be
\beq
\lp \frac{-1-3\vh+\sqrt{1+3\vh}}{\vh}\,, \frac{3}{2} \rp\,,
\eeq
 thus this solution is never an attractor when $\vh$ is positive, but always a saddle point.

\item Kinetic scaling solution with $\Omega_{DBI}=0$. In analogy with the above, this fixed point and its total equation of state are given by the expressions (\ref{uv1}) and (\ref{wuv1}), but now the stability
properties differ because of the different warp factor and potential.
From the eigenvalues
\beq
\lp -\frac{1+3\vh-\sqrt{1+3\vh}}{\vh}\,, -3 \rp\,.
\eeq
we see that this solution is always an attractor.
The difference with  the $m=4$, $n=2$ case is that in that case, the matter scaling solution is a saddle point and the kinetic solution an attractor in the accelerating case $\vh>1$,
while  in the present case, we find that  for all $\vh>0$ the matter scaling solution is a saddle point, while the kinetic scaling solution is an attractor.

\end{itemize}

The fixed points and their stability properties in the two examples considered above are summarised in Table \ref{table}.
\begin{center}
\begin{table}
  \begin{tabular}{| c | c | c | c |}
    \hline
    Fixed point & Stability when & Stability for & $w$ \\
                       &  $m=4,n=2$ &  $m=0,n=-2$ & \\
\hline
    \hline
    (a) Matter domination & Unstable & Unstable & $0$ \\ \hline
   (b) de Sitter solution & Saddle & Saddle & $-1$ \\ \hline
    (c) Kinetic domination & Saddle & Saddle & $1$ \\ \hline
    (d) Matter scaling solution & Attractor iff $0<\vh<1$  & Saddle & $-\frac{\lp 1-\sqrt{1+3\vh}\rp^2}{3\vh}$ \\ \hline
   (d) Kinetic  scaling solution & Attractor iff $\vh>1$  & Attractor & $-\frac{\lp 1-\sqrt{1+3\vh}\rp^2}{3\vh}$ \\ \hline
  \end{tabular}
\caption{Summary of the fixed points in the two examples considered.
\label{table}}
\end{table}
\end{center}

In the dynamical system analyses of DBI cosmologies, scaling solutions have been found in the literature \cite{Guo:2008sz,Martin:2008xw,Burin,Ahn:2009hu,Ahn:2009xd,Copeland:2010jt}. However they
described nonaccelerating expansion with $w_T=0$. The possibility of scaling with $w_T \neq 0$ appears only when a coupling is taken into account (recall our discussion in section \ref{crosscheck}). However, the phenomenological coupling terms considered in \cite{Kaeonikhom:2012xr} did not give rise to new accelerating fixed points, though in some cases the presence of the coupling could modify the stability properties of the fixed points.


\subsection{Numerical solutions}
\label{sec:numerics}

In this section we investigate the system (\ref{sys1}-\ref{dtildegam}) numerically
in order to confirm the results we expect from the  considerations of Section \ref{crosscheck} and the  dynamical system analysis exposed in Section \ref{sec:phase} above. We also  uncover some of the typical details of the evolution as the system converges to its asymptotic state described by the attracting fixed points. Our aim is to construct realistic cosmological scenarios starting from a standard matter dominated
era\footnote{An alternative scaling scenario starting from a scaling matter era will be briefly described in Section \ref{alternative}.}
(for simplicity, as the new effects appear only at late time cosmology, we have omitted the radiation contribution to the early expansion of the universe) and ending in an accelerating era. For purposes of illustration we use an adS  geometry for the warp factor with a quadratic potential for the scalar field. Thus we set $m=4$ and $n=2$ in  what follows.

When we start integrating the equations of motion with matter dominated initial conditions, we generically end in a stage where the DBI field has a significant impact to the dynamics. This was expected since the matter dominated solution is a repeller in all the models at hand (class (a) above). We also generically find the universe ending in an accelerating phase described by the DBI fixed points when $\vh>1$. For the quadratic adS model, we found two of these points:  an attractor and  a saddle point. A typical evolution is such that the universe evolves via the saddle point (\ref{uv1}) into the attractor (\ref{uv2}). Examples are
shown in  figure \ref{plots}, for two  values of the parameter $\vh$. Since the universe typically spends a few e-folds in the saddle point stage and in a realistic case the acceleration has begun only recently, the prediction is that our universe resides (or rather, is reaching) the accelerating scaling saddle point. Interestingly, the equations of state (\ref{wuv1}) and (\ref{wuv2}) for these two physically distinct fixed points coincide, which means that judging from the expansion of the universe alone, they cannot be observationally distinguished. For the fixed point (\ref{uv1}) there is a non-negligible contribution from the disformally coupled dark matter, and for the fixed point (\ref{uv2}) a non-negligible contribution from the kinetic energy of the scalar field\footnote{The reason that they still can have identical expansion rates is that neither dust-like matter nor  the kinetic energy of the scalar field, when suppressed by the relativistic Lorenz factor, contribute to the effective pressure, as seen from Eq.~(\ref{t_eos2}).}.


\begin{figure}
\includegraphics[width=0.5\textwidth]{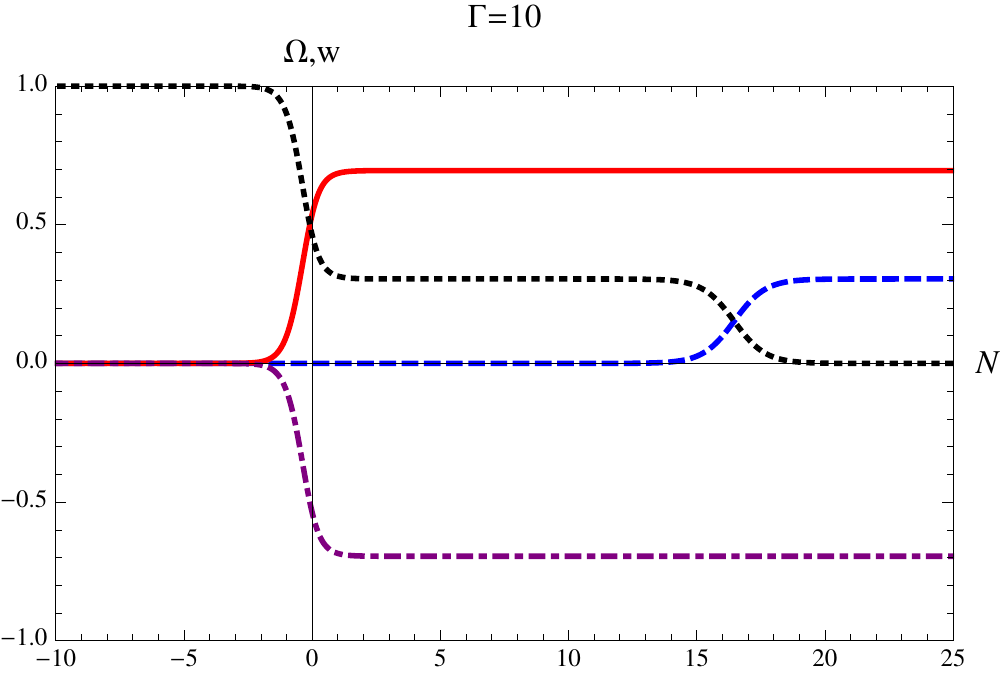}
\includegraphics[width=0.5\textwidth]{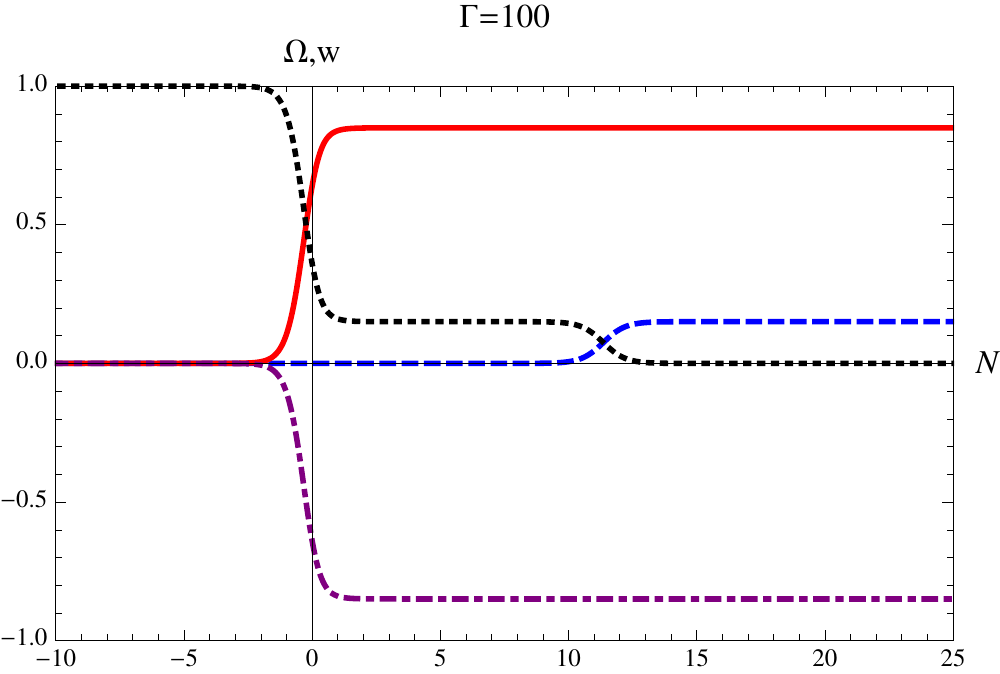}
\caption{The evolution of the fractional energy densities and the total equation of state as functions of the e-folding time $N=\log{a}$ for $\Gamma_0=10$ (left panel) and $\Gamma_0=100$ (right panel). The equation of state is the dash-dotted purple line that settles to its attractor value Eq.~(\ref{wuv1}). The black dotted line is the $\Omega$ for matter
that drops first from the matter-dominated value $\Omega=1$ to the saddle point solution value given by Eq.~(\ref{uv1}) and then to zero as the universe eventually reaches the attractor
described by Eq.~(\ref{uv2}). At the latter transition, the kinetic energy contribution of the field, $x^2$, plotted as the blue dashed line, becomes important. The potential energy contribution $z^2$, plotted as the red solid line, retains its value through the two latter stages.}
\label{plots}
\end{figure}


The various equations of state defined in section \ref{crosscheck} provide another aspect from which to understand the workings of the disformal coupling. In addition to the total equation of state (\ref{t_eos2}), one has the usual definitions for the equations of state of the individual fluids, given in terms of our phase space variables as
\beq
w_{DDM}=0\,, \quad w_{\phi} = \frac{\tgam x^2-z^2}{x^2+z^2}\,.
\eeq
In the presence of the nonminimal coupling however, the scaling of the energy components is defined by the effective equations of state that now can be written as
\beq
w_{DDM}^{eff} = \frac{1}{3}\lb \frac{1}{\tgam}\frac{d\tgam}{dN} \pm \sqrt{\frac{3(1-\tgam^2)}{\vh}} \,z\rb\,, \quad w_\phi^{eff} = w_\phi - \frac{1-x^2-z^2}{x^2+z^2}w_{DDM}^{eff}\,,
\eeq
where the positive sign should be chosen now in the former equation, and the derivative of $\tgam$ is given by Eq.~(\ref{dtildegam}).
The time evolution for these quantities is shown in figure \ref{plotsw}. Because  $\gamma\phi$ grows with time, there is energy transfer from the scalar field to dark matter that makes the latter dilute slower, as is discussed in Sec.~(\ref{crosscheck}). During the scaling era, by definition, $w_T=w_{DDM}^{eff}=w_\phi^{eff}$. Even when this era ends, the coupling continues to slow down the dilution of the DDM energy density, so that 
$w_{DDM}^{eff}$ remains at a constant negative value. In the right panel of figure \ref{plotsw}, we show an example of a case when initially the energy density of the field is not potential-dominated. Then the kinetic scaling era begins shortly after the coupling becomes efficient, and the scaling behaviour never quite takes place. Such initial conditions require the coupling and the kinetic contribution to both become significant around the present epoch, and are thus less generic than the initial conditions that allow some e-folds of scaling.
An interesting detail to observe is that due to the fact that we have set the scalar field evolving as an initial condition, the coupling is effective from early on: in particular, as suggested in Sec.~(\ref{crosscheck}), it forces the energy density of the DBI field to remain constant, \emph{i.e.} $w_\phi^{eff}=-1$ even though $w_\phi>-1$.
This is because the effect of the coupling is to produce an energy flow from dark matter to dark energy, which contributes a very tiny positive $w_{DDM}^{eff}$: when $w_{\phi}^{eff}=-1$,
$w_{DDM}^{eff}=(1+w_{\phi})\rho_\phi/\rho$ as seen from (\ref{weffsc}).

\begin{figure}
\includegraphics[width=0.5\textwidth]{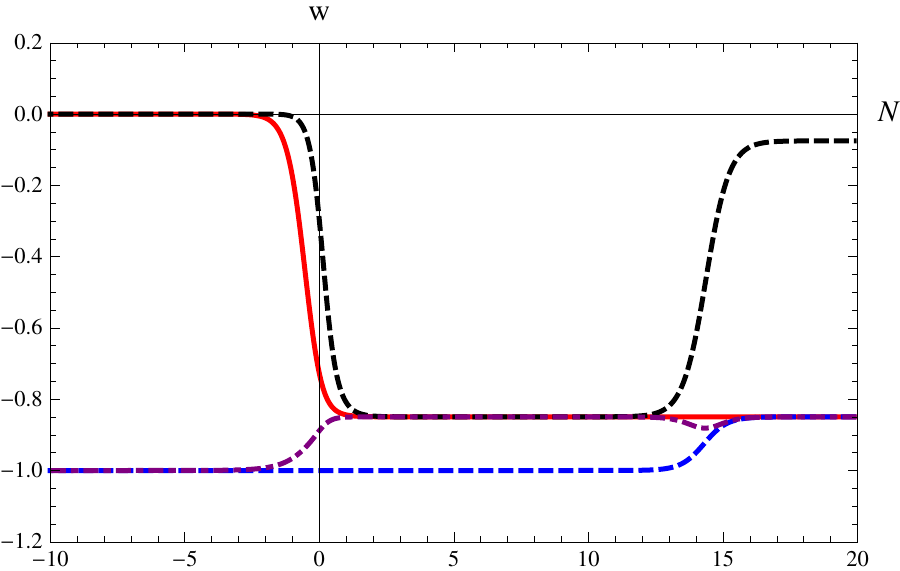}
\includegraphics[width=0.51\textwidth]{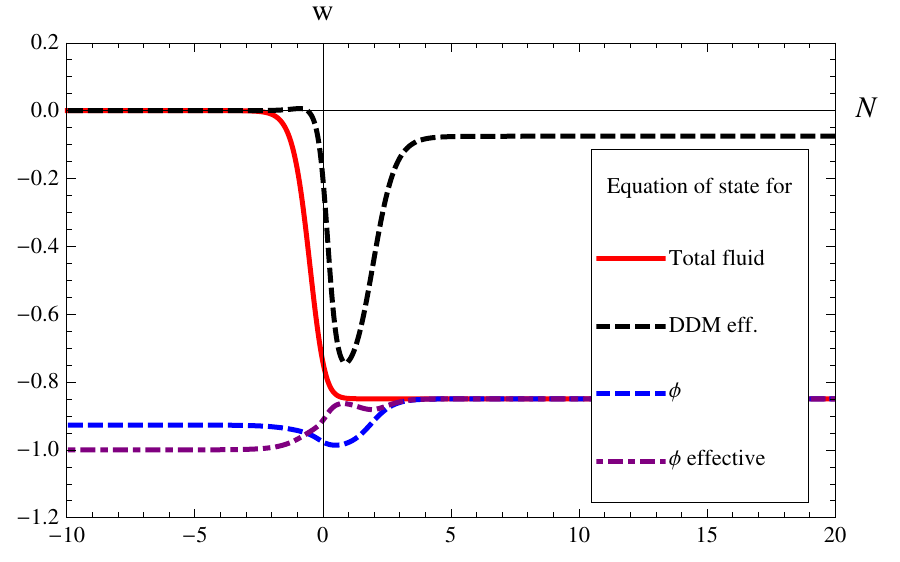}
\caption{The time evolution of the various equations of state as functions of the e-floding time $N=\log{a}$ when $\vh=50$. In the left panel the kinetic energy $x$ is initially small and the $w_\phi=p_\phi/\rho_\phi$ (purple dash-dotted line) as well as the effective equation of state for the field $w^{eff}_\phi$ (blue dashed line) are essentially $w_\phi=-1$ until the coupling begins to modify the dynamics. The effect of the the coupling is to increase the $w_\phi^{eff}$ and to lower the effective equation of state for dark matter $w^{eff}_{DDM}$ (black dotted line) so they both track the total equation of state $w$ (red thick line) during the scaling epoch. When this epoch ends, the dark matter dilutes faster than dark energy, but as seen from the plot, the coupling continues to have an effect on the DDM-component. In the right panel, initial conditions are set such that the kinetic energy $x$ is significant and thus $w_\phi>-1$. In such a case the universe evolves to the kinetic attractor soon after the coupling kicks in, before the scaling solution is reached.}
\label{plotsw}
\end{figure}

To get a better understanding at the dynamics behind this evolution and the role of initial conditions, we plot the variable $x$ and the Lorenz factor $\gamma$ as functions of the scale factor in figure \ref{plots2} for different initial values of $\gamma$. We start with a small $x$ and $z$: for a fixed $\gamma$, the initial value of $z$ determines when we enter into the saddle point, and the initial value of $x$ when into the attractor. We see that the transition from the accelerating fixed point to another occurs when $x$ reaches its critical value given by Eq.~(\ref{uv2}). The more nonrelativistic  $\gamma$ is, the longer this will take. If the brane starts moving very slowly from a virtually non-warped region in  the early universe, after reaching the matter scaling fixed point the universe can stay there for in principle arbitrary number of e-folds before the brane has reached close enough to the tip of the throat to end the matter scaling behaviour. On the other hand, if the initial conditions are relativistic enough the $x$-variable grows with a ``saturated'' rate also during matter dominated epoch and there is no difference in the observational predictions.
In the right panel of figure \ref{plots2}  we see that the scaling of the $\gamma$-factor, which is identical for all initial values during the matter epochs, changes only when the attractor is reached.
The scaling is such that $\gamma\phi \sim a^{-3w_T}$, as expected already from the considerations in Section \ref{crosscheck}.

\begin{figure}
\includegraphics[width=0.5\textwidth]{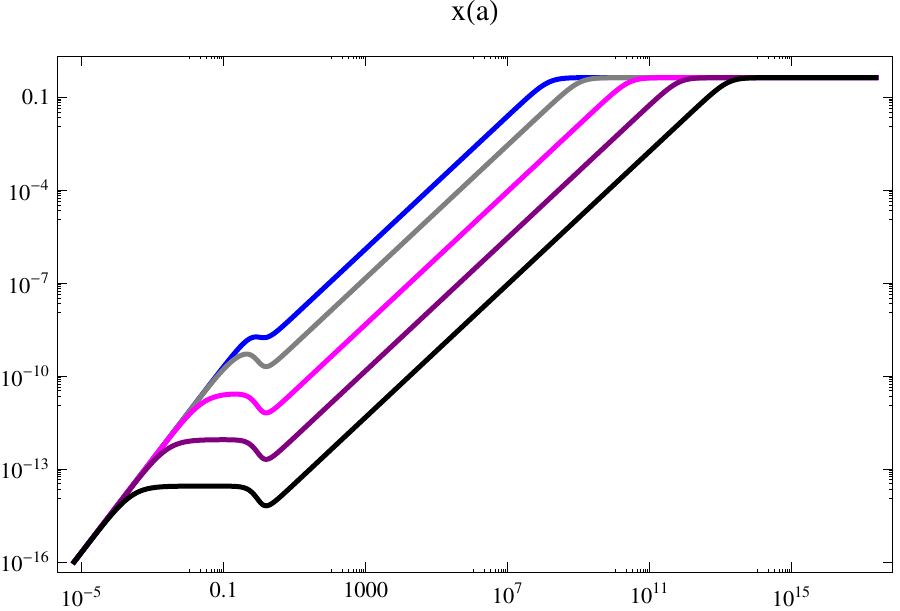}
\includegraphics[width=0.55\textwidth]{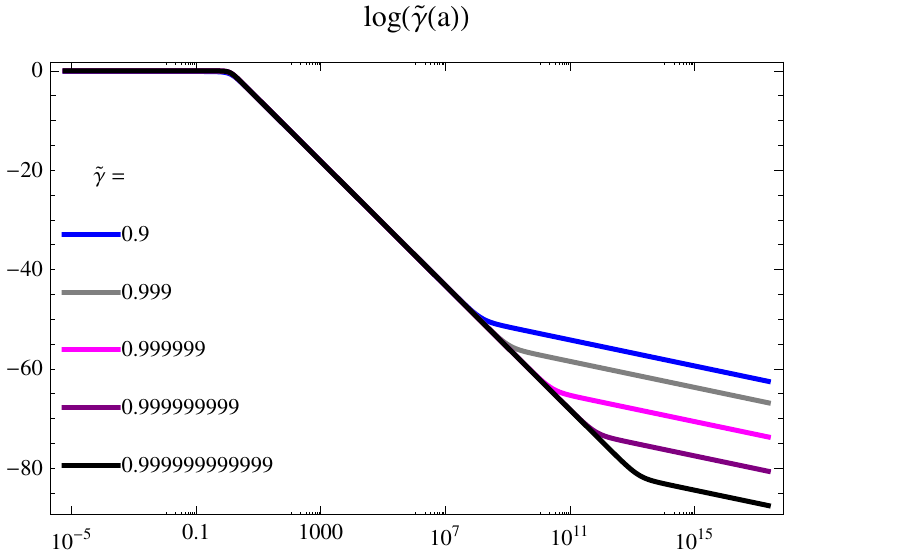}
\caption{The evolution of the ``kinetic term'' $x$ (left panel) and the brane Lorenz factor presented via $\log{\tgam}$ (right panel) as functions of the scale factor $a$ when $\vh=30$. The results are presented for five different initial conditions (set at $a=10^{-12}$) as given in the legend of the right panel. We see that the $x$, initially set to a small value, grows until it reaches the attractor value given by Eq.~(\ref{uv2}). For sufficiently non-relativistic initial condition ($\tgam$ very close to unity), $x$ can be frozen during the matter dominated era but starts growing as the universe enters into the accelerating scaling saddle point solution Eq.~(\ref{uv1}). For sufficiently relativistic initial conditions ($\tgam$ very close to zero) this does not occur. During matter dominated era $\gamma$ is constant, but begins to evolve at a constant rate towards relativistic values $\gamma \rightarrow \infty$ as the accelerating era begins. When the attractor is reached, this rate changes. The rate is given by $\vh$ in such a way that
$\gamma\phi \sim a^{-3w}$ where $w$ is the equation of state in Eq.~(\ref{wuv1}), as expected from considerations in \ref{crosscheck}.}
\label{plots2}
\end{figure}

Finally we check how cosmology depends upon the parameter $\vh$, which is the sole theoretical quantity that controls the evolution. We illustrate this in figure \ref{plots3} by plotting  $x$ and $\Omega$ as functions of the scale factor for $\vh$ of a few different orders of magnitude. In complete agreement with the results of the analytic study in section \ref{sec:phase}, we find that the $\vh=1$ is the dividing value above which the universe accelerates and eventually ends with $\Omega=0$, and below which the universe decelerates forever and $\Omega$ retains a constant finite value.

\begin{figure}
\begin{center}
\includegraphics[width=0.47\columnwidth]{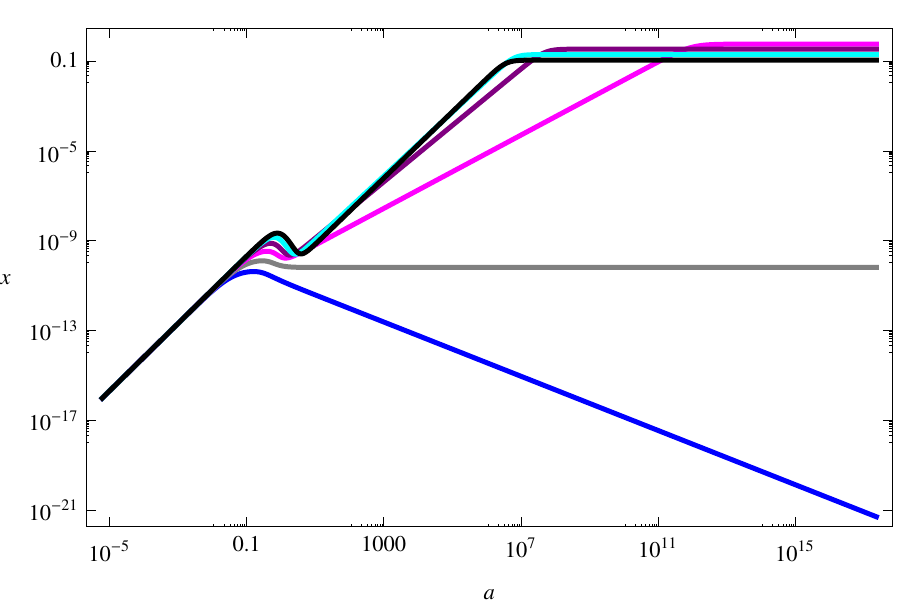}
\includegraphics[width=0.52\columnwidth]{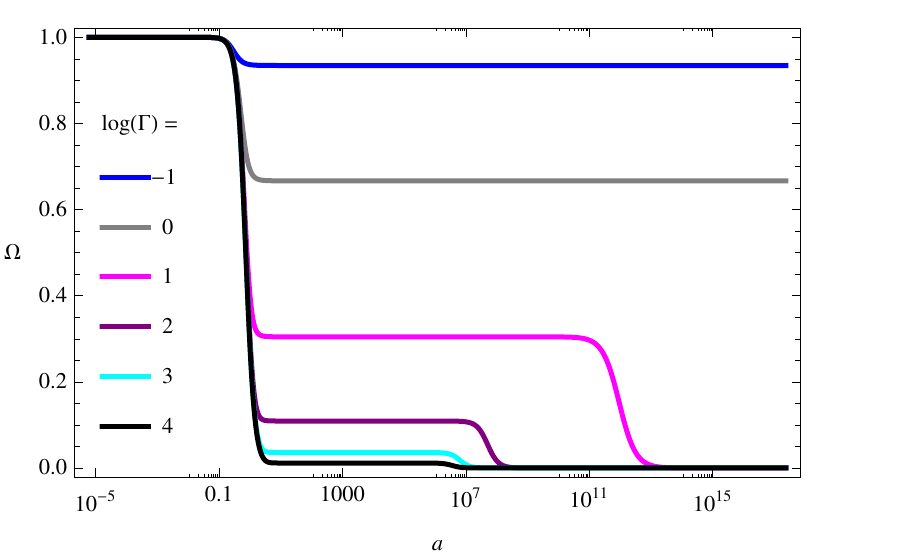}
\caption{The evolution of the ``kinetic term'' $x$ (left panel) and  $\Omega$ (right panel) as functions of the scale factor $a$ for different values of $\vh$ (as given in the legend of the right panel). For $\vh$ larger than unity, the evolution of $x$ is similar to as depicted in figure \ref{plots2}, and $\Omega$ behaves as depicted in figure (\ref{plots}). For the
limiting value $\vh=1$, for which the attractor value of the equation of state is $w=-1/3$, the $x$-term freezes and it occurs that the matter scaling persists. When $\vh<1$, the attractor value of the equation of state is non-accelerating $w>-1/3$ and instead of growing the $x$-term begins to decay when the matter scaling solution is reached. The solution is now an attractor and $\Omega$ remains the constant given in Eq.~(\ref{uv1}).
\label{plots3}}
\end{center}
\end{figure}

\subsection{The vacuum energy scale today}
\label{reality} 

We pause here to discuss the value of the mass and expectation value of the dark energy field today. 
As we now argue, the adS$_{5}$ case with a quadratic potential turns out to be the most appealing model which can provide a tantalising alternative to explain the smallness of dark energy density today. 
As we have seen,   the interesting solutions in this case require $\Gamma_0>1$. This  is given in terms of the 't Hooft parameter $\lambda_{adS}$ and the mass of the field, related to $V_0$. As we have discussed, our approximations are valid so long $\lambda_{adS}\gg1$.
Now, the observed value of the Hubble rate $H$ today is $H \sim 10^{-33}$ eV, corresponding to the 
energy scale of $V_{vac} \sim (10^{-3} {\rm eV})^4$. This can be translated into the condition that $(\kappa m)(\kappa \phi_0) \sim 10^{-60}$.  Since we need $\Gamma_0>1$, the smallest value that the mass of the disformal quintessence field can be is of order $(\kappa m)^2 \gtrsim1/\lambda_{adS}$. Therefore for  reasonable values $\lambda_{adS}\sim 10^{5-10}$,
the mass of the scalar field does not need to be  small. 

On the other hand, the Disformal Dark D-brane is moving towards the tip of the throat at $\phi =0$. Therefore, an observed value of the brane's position today of order $\kappa\phi_0\gtrsim 10^{-60}$ (assuming e.~g.~a mass of the field  around the Planck scale) simply tells us that the brane is exceedingly close to the tip of the throat today! In the very simplest model (using the adS approximation for the warp factor) studied here, the brane reaches the tip of the throat at $\phi=0$ only asymptotically, spending an infinite amount of time getting ever closer with ever slower pace as 
$\gamma \rightarrow \infty$. Therefore a tiny but nonzero value of the field today does not require any coincidence or fine-tuning. We should however point out that the profile of a realistic Klebanov-Strassler throat smoothens to a constant at the tip of the throat, so the warping, though huge, does not grow infinitely. What this may imply is that the universe does not accelerate forever since the field may eventually reach the very tip.

Various comments are in order here. First of all,  we would like to stress that in our scenario we are not interested in  addressing the \emph{cosmological constant problem}. That is, the enormous mismatch between the observed value of the vacuum energy and the expected value of the cosmological constant from theory.
Instead we take another widely studied route in which one assumes that the cosmological constant is either screened or prevented from gravitating by some mechanism, meaning that it is effectively set to zero by these effects\footnote{The cosmological constant will contain the sum of constant vacuum energies of all the fields in the theory, including classical contributions such as possible constant terms in the scalar potential.}. Therefore, we are interested only in  explaining  why the observed vacuum energy is not exactly zero due to the presence of a dynamical scalar field. 
 
 Conceptually, dynamical dark energy models can improve this problem because in principle dark energy can evolve over an enormous range of scales. However, in practise it usually turns out that an incredible amount of fine-tuning is required even from these dynamical models, usually on the mass of the dark energy field, to obtain the correct scale: the scalar mass must be of the order $m \sim 10^{-33}$ eV. 
 One arrives at this conclusion in  standard  quintessence  models, since there the slow roll conditions  generically imply a vacuum expectation value for the scalar today $\phi_0\gtrsim M_{P}$. 
  This is often referred to as the \emph{fine-tuning problem of dark energy}. 

In  our present Dark D-brane scenario, however, the usual slow roll approximation cannot be used, and therefore, such a conclusion does not apply. In fact, in the DBI case, the potential term in the equation of motion for the scalar field is suppressed by the Lorentz  factor (and so is the coupling term in our Disformal Dark D-brane model). Moreover as we have seen, so long as $\Gamma_0>1$ our solutions are valid, even for steep scalar potentials, just as in the DBI inflationary scenario. 
One can see this explicitly  from the analogue of the slow roll conditions in  DBI inflation, which require \cite{AST}:
\beq
\left(\frac{V'}{V} \right)^2 \frac{V h}{ \kappa^2} \gg1 \,.
\eeq
Plugging our power law expressions for the potential and warp factors, this condition becomes
\beq
(\kappa\phi)^{n-m-2}\, \Gamma_0  \gg 1 \,.
\eeq
For the case we have been considering $n-m=-2$, this condition translates into 
\beq
\kappa\phi \ll (\Gamma_0)^{1/4}\,,
\eeq
which is  satisfied for the very small values required to match the vacuum energy today as described above. 
Combining this  with the condition on the dark energy scale today implies 
\beq
\kappa m \gg 10^{-60} (\Gamma_0)^{-1/4}\,,
\eeq
confirming again that the mass of the scalar field does not need to be small, as in standard quintessence models.

\subsection{An alternative scaling scenario}
\label{alternative}

In the example scrutinised in detail above, the universe enters into an accelerating scaling attractor. However, one could also employ the matter scaling solution to alleviate the coincidence problem in the past in such a way that the during the matter dominated era there was a constant contribution from the DBI field energy density to the expansion rate. Then the corresponding equation of state should of course be nearly enough vanishing in order to not spoil the agreement with observations. Indeed, by considering low enough values for $\Gamma_0$ we can make $w_T$ arbitrarily close to zero, and this should work robustly since when $\Gamma_0<1$ the disformal matter scaling solution is an attractor in the phase space. An issue then arises however, is that the phase space trajectory will get stuck to this fixed point and never enter into an accelerating regime. This can be easily changed, but that requires going slightly beyond
the very simplest minimalistic set-up we have focused upon above.

Here we will only briefly explain how the alternative scenario can be naturally realised, and provide a numerical proof of example to confirm it; more detailed study of the alternative scaling scenario is beyond the scope of the present paper. In this scenario, the moving brane has been sliding towards the throat through the history of our universe, resulting in a slightly modified effective behaviour of dark matter and a scaling of the energy density that is identical to that of dark matter. This energy density will become a constant and our universe will begin to accelerate as  a consequence of the moving brane reaching deep enough into the throat where the geometry and the effective potential for the DBI field may need to be modified.

When in the disformal attractor, the scalar field is rolling down the potential, in other words, the moving brane is approaching the tip of the throat and the Lorentz boost factor is increasing towards infinity. At some point, as is well known, the adS approximation  $h \sim \phi^{-4}$ will break down. In the KS geometry the warp factor will become  roughly constant very close to the tip.
On the other hand, the quadratic approximation for the potential may also receive significant corrections. What this generically seems to imply for the dynamics is that the de Sitter solution becomes an attractor.
Thus, as corrections become important the universe will begin to accelerate. Just to illustrate such a case, we modify the potential in such a way that at a suitable point the quadratic slope changes, in the examples shown in figure \ref{plots4}, to either linear or to quartic. This is of course a crude way to take
into account the possible corrections to the simplest possible set-up, and it should be studied in more generality whether the outcome we find here persists in more realistic cases. Investigations of dynamics with more complicated potentials and warp factors are however outside the scope of the present study.

\begin{figure}
\begin{center}
\includegraphics[width=0.49\columnwidth]{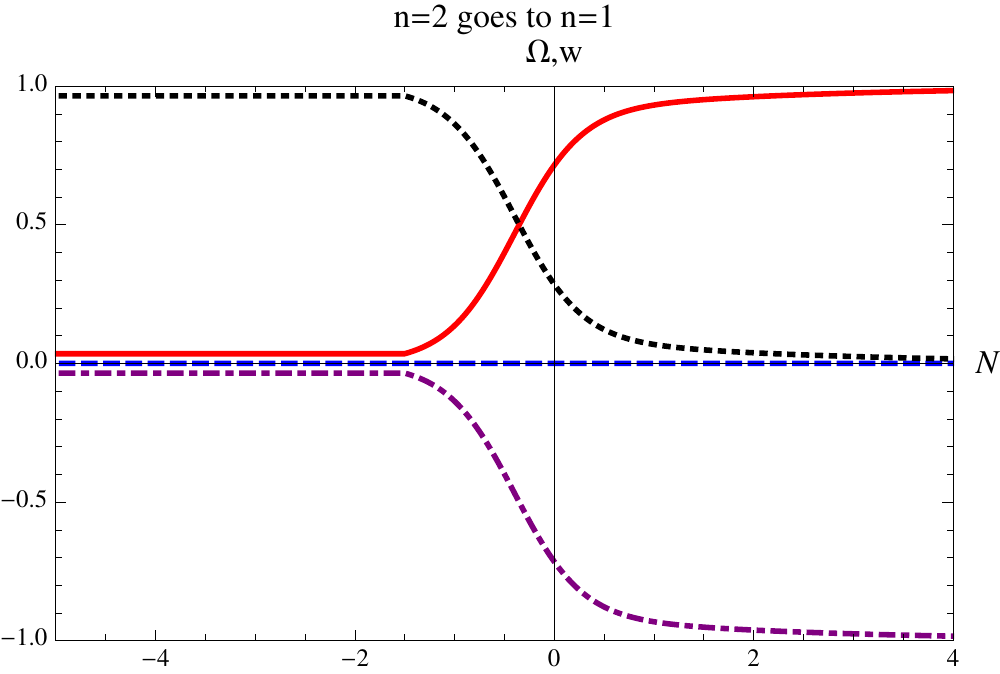}
\includegraphics[width=0.49\columnwidth]{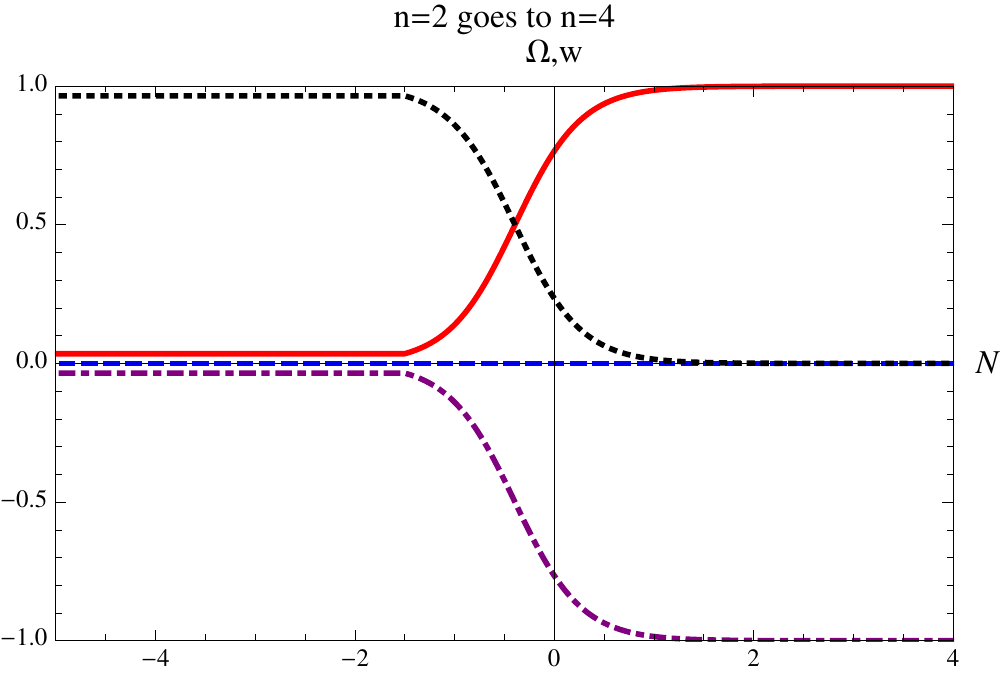}
\caption{The cosmological evolution in the alternative scaling scenario of Section \ref{alternative} when the potential becomes linear (left panel) and quartic (right panel) for small field
values. In both cases, we have set $\Gamma_0=0.05$. The equation of state is the dash-dotted purple line that begins from its slightly negative attractor value Eq.(\ref{wuv1}) and settles to $w=-1$. The black dotted line is the $\Omega$ for matter that drops from the disformal 
scaling value given by Eq.~(\ref{uv1}) to zero as the field domination begins. The potential energy contribution $z^2$, plotted as the red solid line, becomes dominant as the de Sitter stage is reached. The kinetic energy contribution of the field, $x^2$, plotted as the blue dashed line, remains negligible in this scenario. The only difference between the examples is that in the case of asymptotically linear potential, the transition to de Sitter expansion is slightly smoother.
\label{plots4}}
\end{center}
\end{figure}


\section{Conclusions}
\label{sec:conclusions}


In this article we have proposed and explored the cosmological implications of an intuitive geometric picture in string theory compactifications, where the complete dark sector of the universe may be due to a ``hidden sector'' D-brane (or stack of D-branes). In this picture, dark energy emerges from the oscillations of the open strings transverse to the brane, dark matter is due the massive oscillations of the open strings along the brane (DIMPs), and possible dark radiation is due to the massless open string oscillations along the brane.

Dark energy is thus associated to the brane's motion in the compact six dimensional space, which in the four-dimensional theory is described by a scalar field with non-standard (DBI) kinetic terms. 
We allow the moving brane to carry matter and therefore dark matter (and/or dark radiation) is associated with these matter fields living on the brane, which couple in a precise way to dark energy. In particular, they couple via a disformal relation \cite{Bekenstein}, thus giving an explicit realisation of this relation and a new framework to construct disformally coupled quintessence models in string theory\footnote{Phenomenological models of disformal quintessence have been explored recently in a handful of papers: to highlight one interesting result, the disformal coupling was discovered to feature a new kind of efficient and largely model-independent screening mechanism \cite{Koivisto:2012za}.
Moreover, an stringy inspired coupled quintessence model was proposed in \cite{CPT}.}.

As has been shown in the literature, the coincidence problem can be relaxed in coupled models of DM and DE, in as much as an interchange ({\it i.e.}~coupling) between these two forms of energy can explain why they are of the same order today (see \cite{Ukranian} for a review). Moreover, whereas new forces between DE and normal matter are heavily constrained by observations ({\it e.g.}~in the solar system and gravitational experiments on Earth), this is not the case for DM.
A resolution of the ``cosmic coincidence'' problem then implies  that energy densities of DE and DM dilute with the same rate for a significant period of the universe's expansion.
Therefore the Disformal Dark D-brane scenario we propose can alleviate this problem and can also explain why a coupling among DE and standard model fields is suppressed, since the Disformal Dark D-brane is a hidden brane and thus ``dark'' by construction.

We have considered the simplest Dark D-brane model in terms of a probe Dark D3-brane moving in the throat of a warped compactification in type IIB string theory, and have explored the resulting homogeneous cosmological evolution.
We studied the system of equations in terms of a set of dimensionless and bounded phase space variables in order to conveniently undertake a dynamical system analysis. We focused on a power law form for the warp factor and the  potential and for this case  derived four classes of fixed points, (a, b, c, d), from which only one is independent of the values of the parameters (class (a)). The other fixed points depend upon the value of the exponents in the power law expressions for the warp factor and the potential, namely $(n-m)$ (see Sec.~\ref{sec:phase}).
 Furthermore the most interesting class of fixed points, class (d), which contains scaling accelerating solutions, turned out to depend upon a single dimensionless parameter $\vh$, the ratio of the scales of the potential and of the warp factor. This clarifies how the cosmological dynamics are invariant under a given class of rescalings of the fields and the parameters.

Within this interesting class of fixed points, we have studied explicitly two representative cases: Firstly, we considered a Disformal Dark D-brane moving in an adS$_5$ throat, which can be seen as an approximation to a mid-throat region in a KS geometry with a quadratic potential. Secondly, a Disformal Dark D-brane moving in a geometry with a constant warp factor, which can arise as a very near-tip KS region or simply as an unwrapped region in a large volume scenario, with an inverse power law potential.
In the ultra-relativistic limit, we found two different types of disformal scaling fixed points for the two cases (see Table \ref{table}). At one of these points, the (disformally coupled) dark matter contributes a constant fraction to the expansion rate (the matter scaling solutions in Sec.~\ref{sec:phase}).
At the other, the matter sources contribute negligibly to the energy density (the kinetic scaling solutions), and yet this fixed point exhibits precisely the same total equation of state as the previous fixed point.
The stability properties of these two types of fixed points for the adS$_5$ and constant warp factor cases depends on the value of $\vh$, and differs for both cases as follows (see Table \ref{table}):
 In the adS$_5$ case for $\vh>1$, the matter scaling fixed point is an accelerating saddle point, while the kinetic scaling solution is an accelerating stable fixed point.
 Instead, for the constant warp factor with inverse power low potential, the matter scaling solution is a saddle point while the kinetic fixed point is an attractor for all positive values of $\vh>0$.

 Finally, we confirmed numerically all of the results which were expected from the analytic and dynamical system analysis in the adS$_{5}$ case in Sec.~\ref{sec:numerics}.
In summary, we found that a generic background expansion is such that presently the universe is undergoing an accelerating scaling expansion, but asymptotically becomes devoid of matter while still expanding with the identical accelerating rate. We also presented an alternative application of the new scaling solution, where a de Sitter solution is reached after a scaling matter era.

As we discussed in Section \ref{reality}, the adS$_{5}$ case suggests a compelling geometrical explanation for the smallness of DE today. Since the value of the quintessence potential today is required to be very small,  $V_{vac}\sim (10^{-3}{\rm eV})^4$, in terms of a quadratic potential, this implies that $(m\,\phi_0) \sim 10^{-60} M_P^2$. Since the brane is moving towards the tip of the throat at exactly $\phi=0$, the smallness of the vacuum energy today can be simply translated into the fact that today, the Disformal Dark D-brane is very close to the tip of the throat, \emph{i.e.}~at a field value of $\phi_0\gtrsim 10^{-60}M_P$. 
Therefore, the mass of the DBI quintessence field does not need to be fine tuned to be extremely small as in standard slow roll quintessence. 

Our Disformal Dark D-brane proposal has several natural advantages with respect to early inflation DBI models. Firstly, because it describes late time acceleration of the universe, there is no reason for the moving D-brane to be empty; whereas in early universe inflation models we can expect the matter to have diluted away. Secondly, several standard problems of DBI inflation, or early universe DBI acceleration, do not arise, such as the Lorentz factor becoming too large and thus causing problems with back reaction \cite{EGTZ} and overly large non-Gaussianites. 
Finally, as we already mentioned, there is in principle no reason for the D-brane position to have reached its minimum, as long as it does not cause problems for cosmology. Therefore, while we have not yet produced a completely realistic coupled DM/DE model from the Disformal Dark D-brane scenario, it seems a very promising avenue to explore new ways to test higher dimensional theories, in particular, string theory.

There are clearly several steps that need to be carried out in order to assess whether the Dark D-brane picture can indeed explain the observed dark sector of our universe.
Firstly, from a phenomenological point of view, in order to check the cosmological viability of the model the next step is to
explore the physics of structure formation, since the disformally coupled nature of dark matter can result in new phenomenology at the linear and nonlinear levels. This may allow us to constrain the models with high precision and to distinguish them from other alternative explanations of the dark sector. Preliminary studies of cosmologies with  disformal couplings have uncovered a very rich structure in the disformally coupled perturbation equations for clustering of dark matter (compared to the conformally coupled ones) \cite{Bettoni:2011fs,Bettoni:2012xv,Koivisto:2012za,Zumalacarregui:2012us}, but they have not been implemented in the type of models  presented here.  The massless modes residing upon the Dark D-brane would also be worth investigating in detail as they will contribute possible dark radiation to our four-dimensional cosmology. What again distinguishes this proposal from other models of the dark sector is that the radiation will be disformally coupled, entailing novel phenomenology that can provide the possibility to test Disformal Dark D-brane cosmology with experimental data.

From the theoretical point of view, our proposal requires a more rigorous construction, possibly including the observable sector 
as well as the implications (if any) for other stabilised moduli,
and a compelling argument for the hidden sector brane to be moving today,  although this may require a better understanding of the reheating mechanism in D-brane inflation. 

One immediate  question which arises in our present model is the choice of a mass term potential. In principle one could argue for possible symmetries which justify such a choice \cite{Silverstein:2003hf,AST}. However, one may expect that other terms should generically appear \cite{Baumann:2010sx}. An investigation of solutions for more general potentials and warp factors is under current investigation. 
A further avenue to pursue would be the possible unification of the early and late time acceleration in  a string theory set up. For example, an interplay between open and closed string moduli seems to be a natural possibility, namely early time acceleration originated from the closed string sector, while late time acceleration originated from an open string sector given by our  Dark D-brane world scenario.
We leave the study of these and other interesting ideas for  future investigation.

The history of physics is a testimony to the elegance of Nature, as with the passing of time ever more apparently unrelated phenomena are found to be unified within a deeper structure. String theory suggests that quantum field theory and gravity may be unified as distinct phenomena arising from the oscillations of a single object, the string,
in a higher dimensional spacetime. In line with this principle, the results of the current work suggest that the ``great unknowns'' of cosmology, namely dark energy, dark matter and possibly dark radiation, may all be but manifestations of different aspects of D-brane fluctuations. It remains to be seen whether this picture is supported by the precision data of cosmological large scale structure, and whether such data may guide us towards constructing more detailed scenarios, providing new avenues for experimental tests of string phenomenology.

\section*{Acknowledgements}
TK was supported by the Research Council of Norway and DW by an STFC studentship. We would like to thank Carsten van de Bruck, Cliff Burgess, Chris Clarkson, Ruth Gregory, David Mota, Johannes N\"oller, Roy Maartens, Gianmassimo Tasinato, David Wands and Miguel Zumalac\'{a}rregui for disformal discussions.

\begin{appendix}

\section{Disformalities} \label{appendixGen}

\begin{itemize}
  \item Metric, inverse, determinant and connection:

  In terms of general functions $C=C(\phi,X)$ and $D=D(\phi,X)$ where $X = g^{\mu\nu}\phi_{,\mu}\phi_{,\nu}$ is the kinetic term of the scalar, the disformal metric and its inverse are given respectively by
  \beq
  \bar{g}_{\mu\nu}= C(\phi, X)g_{\mu\nu}+ D(\phi,X)\partial_\mu \phi \partial_\nu\phi, \hspace{1cm}\bar{g}^{\mu\nu}=\frac{1}{C}\left[g^{\mu\nu}-\frac{D\partial^\mu\phi \partial^\nu\phi}{C+D(\partial\phi)^2}\right].
  \eeq

  The determinant $\sqrt{-\bar{g}}$ of the disformal metric may be expressed in terms of the determinant $\sqrt{-g}$ of the background spacetime as
  \beq\label{disformaldet}
  \sqrt{-\bar{g}}=C^2\sqrt{1+\frac{D}{C}(\partial\phi)^2}\sqrt{-g}.
  \eeq

  Finally, for the case in which $C=C(\phi)$ and $D=D(\phi)$, the disformal Levi-Civita connection takes the compact form\footnote{See \cite{Zumalacarregui:2012us} for the general case.}
  \beq\label{AppConnection}
\bar{\Gamma}^\mu_{\alpha\beta}=\Gamma^\mu_{\alpha\beta} + \frac{C'}{C}\delta^\mu\!_{(\alpha} \partial_{\beta)}\phi +\frac{D}{(C+D(\partial\phi)^2)}\partial^\mu\phi\left(\nabla_\alpha\nabla_\beta\phi-\frac{C'}{2D}g_{\alpha\beta}+\left(\frac{D'}{2D}-\frac{C'}{C}\right)\partial_\alpha\phi\partial_\beta\phi\right).
\eeq
  \item Lengths and angles:

  The norm of a vector in takes the form
  \beq
  \bar{g}_{\mu\nu}x^\mu x^\nu = C \,x^2 + D (\partial\phi \cdot x)^2,
  \eeq
  where $x^2 \equiv x\cdot x = g_{\mu\nu}x^{\mu}x^{\nu}$ is the norm in the background spacetime and $\partial\phi \cdot x = \partial_\mu\phi \, x^\mu$. The inverted relation is
  \beq
  x^2 = \frac{x\,\hat{\cdot}\,x}{C}  - \frac{D}{C} (\partial\phi \cdot x)^2.
  \eeq

  The angle between two vectors $x$ and $y$ is given by
  \beq
  \cos\bar{\theta} = \frac{x\cdot y +\frac{D}{C}(\partial\phi \cdot x)(\partial\phi \cdot y)}{\mid x\mid\mid y\mid\sqrt{1+\frac{D}{C x^2}(\partial\phi \cdot x)^2}\sqrt{1+\frac{D}{C y^2}(\partial\phi \cdot y)^2}},
  \eeq
  where $\mid x\mid = \sqrt{x\cdot x}$. The inverted relation is
  \beq
  \cos{\theta} = \frac{x\,\hat{\cdot}\, y - D(\partial\phi \cdot x)(\partial\phi \cdot y)}{\mid \hat{x}\mid\mid \hat{y}\mid\sqrt{1-\frac{D}{\hat{x}^2}(\partial\phi \cdot x)^2}\sqrt{1-\frac{D}{\hat{y}^2}(\partial\phi \cdot y)^2}}\,,
  \eeq
  from which it is obvious that the conformal relation has nothing to do with the distortion of angles.

  \item The coupling term:

The non-conservation coupling term takes the form \cite{Koivisto:2012za}
\beq\label{AppQ}
Q = \frac{C'}{2C}T-\nabla_\mu \lp \frac{D}{C} \partial_\nu\phi T^{\mu\nu}\rp
+ \frac{D'}{2C}\partial_\mu\phi\partial_\nu\phi T^{\mu\nu}\,,
\eeq
which in an FRW background becomes (the subscript $0$ denotes the background value)
\beq \label{q1}
Q_0 = \frac{\rho}{C}\lb\frac{1}{2}C'-\frac{C'D}{C}\dot{\phi}^2 + 3DH\dot{\phi}+\frac{1}{2}D'\dot{\phi}^2+D\lp \dot{\phi}\frac{\dot{\rho}}{\rho}+\ddot{\phi}\rp\rb\,.
\eeq
\end{itemize}

\section{Exponential warps and potentials}
\label{expos}

Suppose that the potential and the warp factor have the form,
\beq
V \sim  e^{\kappa\lambda\phi}\,, \quad h \sim e^{-\kappa\mu\phi}\,.
\eeq
respectively, where $\la$ and $\mu$ are constants.
The dynamics of the phase space is described by the equations (\ref{dxdn}-\ref{dgdn}). The dynamics is oblivious to the energy scales of these functions which we need not thus specify.
An exponential warp could perhaps be motivated in some brane world scenarios, but is not supported by the extra-dimensional
geometries we discussed in Section \ref{sec:geometry}.
In the following we shall briefly list the fixed points of the system and their properties.
\begin{itemize}
\item Matter dominated solution: $x=z=0, \,\, \Omega=1,\,\, w=0$.
This solution exists regardless of $\gamma$ and is a repellor.
\item Potential dominated solution $x=0,\,\, z=1,\,\,\Omega=0,\,\, \tgam=1,\,\, w=-1$. This solution exists when $\la=0$ and is then an attractor.
\item Kinetic dominated solution: $x=\pm1,\,\, z=0,\,\,\Omega=0,\,\, w=\tgam$. This solution exists regardless of $\tgam$. In the limit $\tgam=0$ it is a saddle point, in the limit $\tgam=1$ the downhill branch is an attractor given $\la<-\sqrt{6}$ and $\mu >  2\sqrt{6}$, and symmetrically, the uphill branch is an attractor given $\la>\sqrt{6}$ and $\mu <  -2\sqrt{6}$.
\item Kinetic scaling solution: $x=-\mu/2\sqrt{6},\,\, z=0,\,\,\Omega=1-\mu^2/24,\,\, w=0,\,\, \tgam=1$. This solution exists when $|\mu|<2\sqrt{6}$ but is never stable.
\item Field dominated solution:  $x=\la/6,\,\, z=\sqrt{1-\la^2},\,\,\Omega=0,\,\, \tgam=1,\,\, w=-1+\la^2/3$. This solution exists when $0<|\la|\le\sqrt{6}$ and $\la(\la+4\mu)\ge 12$. Further,
it accelerates when $0<|\la|<\sqrt{2}$.
\end{itemize}
Thus no new scaling attractors are found in the system. The potential accelerating solutions correspond to those present for uncoupled exponential quintessence as well. However, the coupling can modify both the details of the cosmological evolution and that could be used to place constraints on $\mu$.

\section{An alternative autonomous system} \label{auto}

Here we rewrite the set of cosmological equations of Section as a first order autonomous system using alternative variables to span the phase space from Section \ref{sec:phase}.
We have experimented with several different choices of variables and their combinations in order to find the most convenient ones for our purposes. The set $x$ and $z$, supplemented with $\tgam$ turned out to be the most practical and lead to simplest evolution equations. 
An alternative formulation can be better suited for {\it e.g.}~numerical studies of some aspects. 
In general also the fixed points of a dynamical system can depend upon the variables used to formulate it. For these reasons it might be useful to present an alternative formulation here. The variables employed here correspond to those used in Ref. \cite{Guo:2008sz}.

Let us define the dimensionless variables
\beq
\x=\frac{\kappa}{\sqrt{3}H}\sqrt{\frac{\gamma}{h}}\,, \quad y = \frac{\kappa\dot{\phi}\sqrt{\ga}}{H}\,, \quad z = \frac{\kappa\sqrt{V}}{\sqrt{3}H}\,, \quad \Omega=\frac{\kappa^2\rho}{3H^2}\,.
\eeq
In comparison to variables used in Section \ref{sec:phase}, here $\chi=x/\sqrt{1-\tgam}$ and $y=\pm \sqrt{3(1+\tgam)}x$.
Only three of the four variables are independent due to the Friedmann constraint. We choose to eliminate $\Omega$ by
\beq \label{om}
\Omega=1-\lp 1-\gamma^{-1}\rp x^2-z^2\,.
\eeq
The evolution equations for the remaining three variables can be written as
\bea
\frac{d\chi }{dN} & = &
\frac{1}{24 \gamma ^2 \left(3 \left(\gamma ^3-\gamma +1\right) \chi ^3-\gamma
   \chi  \left(\gamma ^2 y^2+3 z^2-3\right)\right)}
   \Bigg[
2 \gamma ^{13/2} \mu  y^5+12 \gamma ^5 y^4 \nonumber \\
& - & 3 \gamma ^{7/2} y^3 \left(\left(4 \gamma ^3-\gamma +3\right) \mu  \chi ^2-\gamma
   \mu +\gamma  z^2 (4 \lambda +\mu )\right) \nonumber \\
  & + & 36 \gamma ^3 \chi ^2 y^2 \left(-\gamma  \chi ^2+\chi ^2+\gamma ^2
   \left(z^2-2\right)\right) \nonumber \\
   & + & 9 \gamma ^{3/2} \chi ^2 y \left(\left(\gamma  \left(\gamma  \left(2 \gamma ^3-\gamma
   +3\right)-4\right)+4\right) \mu  \chi ^2+4 \gamma ^3 \lambda  z^2+\left(\gamma ^2-4\right) \gamma  \mu
   \left(z^2-1\right)\right) \nonumber \\
   & - & 108 \chi ^2 \left(-\gamma  \chi ^2+\chi ^2+\gamma ^2 \left(z^2-1\right)\right) \left(\left(\gamma
   ^3-\gamma +1\right) \chi ^2+\gamma -\gamma  z^2\right)\Bigg]\,,  \\
\frac{dy }{dN} & = &
\frac{1}{24 \gamma ^2 \chi ^2 \left(3 \left(\gamma ^3-\gamma +1\right) \chi ^2-\gamma  \left(\gamma ^2
   y^2+3 z^2-3\right)\right)}
\Bigg[2 \gamma ^{13/2} \mu  y^6+12 \gamma ^5 y^5
\nonumber \\
& - & 3 \gamma ^{7/2} y^4 \left(\left(4 \gamma ^3-5 \gamma +3\right) \mu  \chi ^2-\gamma
    \mu +\gamma  z^2 (4 \lambda +\mu )\right) \nonumber \\
   & + & 36 \gamma ^3 \chi ^2 y^3 \left(-\gamma  \chi ^2+\chi ^2+\gamma ^2
   \left(z^2-2\right)+2\right) \nonumber \\
   &+ &9 \gamma ^{3/2} \chi ^2 y^2 \Big(\left(\gamma  \left(\gamma  \left(2 \gamma ^3-9 \gamma
   +3\right)+6\right)-2\right) \mu  \chi ^2 \nonumber \\
  & +&  4 \gamma  \left(\gamma ^2-2\right) \lambda  z^2+\gamma  \left(\gamma ^2-6\right) \mu
   \left(z^2-1\right)\Big) \nonumber \\
  & + & 108 \chi ^2 y \left((\gamma -1) \Big(\gamma ^3-\gamma +1\right) \chi ^4
  +\gamma  \chi ^2 \Big(\gamma
   \left(\gamma ^3-3 \gamma +2\right)  \nonumber \\
  & + &\left(\gamma  \left(-\gamma ^3+\gamma -2\right)+1\right) z^2-1\Big)+\gamma ^3
   \left(z^2-1\right)^2\Big) \nonumber \\
   & + & 54 \gamma ^{3/2} \chi ^4 \left(\left(2 \gamma ^3-5 \gamma +3\right) \mu  \chi ^2-\gamma  \mu +\gamma
    z^2 (4 \lambda +\mu )\right)\Bigg]\,, \\
    \frac{dz }{dN} & = &
-\frac{z \left(-3 \gamma  \chi ^2+3 \chi ^2+\gamma ^{3/2} \lambda  y+3 \gamma ^2 \left(z^2-1\right)\right)}{2 \gamma ^2}\,.
\eea
Here $\gamma$ should be understood as a shorthand notation for
\beq
\gamma \equiv \frac{\chi}{\sqrt{\chi^2-\frac{1}{3}y^2}}\,.
\eeq
Given $\lambda$ and $\mu$, the system is closed. If we consider the forms given in (\ref{forms}), then $\lambda$ and $\mu$ in the above system can be replaced by
\beq
\mu=m\lb \frac{\ga z^2}{\vh \x ^2}\rb^\frac{1}{m-n}\,, \quad \label{mpres}
\la=-n\lb \frac{\ga z^2}{\vh \x ^2}\rb^\frac{1}{m-n}\,.
\eeq
For completeness, in general they are determined as
\bea
\frac{d\mu}{dN}   & = & \Gamma_\mu \frac{y}{\sqrt{\ga}} \,, \quad \Gamma_\mu = \frac{{h'}^2-h''h}{\kappa^2 h^2}\,, \\
\frac{d\la}{dN} & = & \Gamma_\lambda \frac{y}{\sqrt{\ga}}\,, \quad \Gamma_\la = \frac{{V'}^2-V''V}{\kappa^2 V^2}\,.
\eea

\end{appendix}


\providecommand{\href}[2]{#2}\begingroup\raggedright\endgroup

\end{document}